\documentclass[10pt, conference, letterpaper]{IEEEtran}

\usepackage[noadjust]{cite}

\usepackage{subfigure}
\usepackage{amsmath}
\usepackage{amsthm}
\usepackage{graphicx} 
\usepackage{caption}
\usepackage{graphicx,subfigure}
\usepackage{multicol} 
\usepackage{graphicx} 
\usepackage{amssymb}
\usepackage{float} 
\usepackage{wrapfig} 
\usepackage{booktabs}
\usepackage{lipsum} 
\usepackage{indentfirst}
\usepackage{bm}
\usepackage{color}
\usepackage{amsmath}
\usepackage{enumerate}
\usepackage{extarrows}
\usepackage{url}

\usepackage{graphicx}
\usepackage{epstopdf}
\usepackage{multirow}
\usepackage{tablefootnote}
\usepackage[flushleft]{threeparttable}
\usepackage{slashbox}
\usepackage{array}
\usepackage{amsmath}
\usepackage{enumitem}
\usepackage{framed}

\usepackage{tcolorbox}

\usepackage[ruled,linesnumbered,vlined]{algorithm2e}  
\usepackage{framed}

\IEEEcompsocitemizethanks
\IEEEcompsocthanksitem
\IEEEoverridecommandlockouts

\usepackage{lscape}

\newtheorem{problem}{\textbf{Problem}}

\newtheorem{lemma}{\textbf{Lemma}}

\newtheorem{proposition}{\textbf{Proposition}}

\newtheorem{theorem}{\textbf{Theorem}}

\newtheorem{question}{Question}
\newtheorem{auction}{Auction}

\definecolor{darkgreen}{rgb}{0.13, 0.55, 0.13}

\def\AoI{A}

\def\AoIthr{\AoI^{\textit{thr}}}

\def\rdmSP{\sigma}
\def\rdmPoI{\omega}

\def\bidSP{s}
\def\bidPoI{p}

\def\proc{r}

\def\status{h^{\textit{PoI}}}
\def\Status{h^{\textit{Plat}}}

\def\V{V}

\begin{document}

\title{Taming Time-Varying Information Asymmetry in Fresh Status Acquisition}

\author{
	\IEEEauthorblockN{
		Zhiyuan Wang\IEEEauthorrefmark{1},
		Lin Gao\IEEEauthorrefmark{2}\IEEEauthorrefmark{4}, and
		Jianwei Huang\IEEEauthorrefmark{3}\IEEEauthorrefmark{4}
		\thanks{
			This work is supported by the Shenzhen Institute of Artificial Intelligence and Robotics for Society, and the Presidential Fund from the Chinese University of Hong Kong, Shenzhen.			
			This work is supported by the National Natural Science
			Foundation of China (Grant No. 61972113), 
			Shenzhen Science and Technology Program (Grant No. JCYJ20190806112215116, JCYJ20180306171800589, and KQTD20190929172545139), and Guangdong Science and Technology
			Planning Project under Grant 2018B030322004. 			
		}
	}
	
	\IEEEauthorblockA{\IEEEauthorrefmark{1}Department of Computer Science and Engineering, The Chinese University of Hong Kong}
	\IEEEauthorblockA{\IEEEauthorrefmark{2}School of Electronics and Information Engineering, Harbin Institute of Technology, Shenzhen}
	\IEEEauthorblockA{\IEEEauthorrefmark{3}School of Science and Engineering, The Chinese University of Hong Kong, Shenzhen}
	\IEEEauthorblockA{\IEEEauthorrefmark{4}Shenzhen Institute of Artificial Intelligence and Robotics for Society}
}

\maketitle

\IEEEcompsocitemizethanks
\IEEEcompsocthanksitem
\IEEEoverridecommandlockouts

\begin{abstract}
	Many online platforms are providing valuable real-time contents (e.g., traffic) by continuously acquiring the status of different Points of Interest (PoIs).
	In status acquisition, it is challenging to determine how frequently a PoI should upload its status to a platform, since they are self-interested with private and possibly time-varying preferences.
	This paper considers a general multi-period status acquisition system, aiming to maximize the aggregate social welfare and ensure the platform freshness.
	The freshness is measured by a metric termed age of information.
	For this goal, we devise a long-term decomposition (LtD) mechanism to resolve the time-varying information asymmetry.
	The key idea is to construct a virtual social welfare that only depends on the current private information, and then decompose the per-period operation into multiple distributed bidding problems for the PoIs and platforms.
	The LtD mechanism enables the platforms to achieve a tunable trade-off between payoff maximization and freshness conditions. 
	Moreover, the LtD mechanism retains the same social performance compared to the benchmark with symmetric information and asymptotically ensures the platform freshness conditions.
	Numerical results based on real-world data show that when the platforms pay more attention to payoff maximization, each PoI still obtains a non-negative payoff in the long-term.
\end{abstract}

\section{Introduction}
\subsection{Background and Motivation}
We have witnessed a growing popularity of online content platforms, which provide valuable real-time contents related to people's daily life.
The platforms needs to update its contents based on the real-time status of different Points of Interest (PoIs).
For example, some platforms (e.g., Waze \cite{waze} and MapFactor \cite{MapFactor}) rely on the location and trajectory reports from mobile users to acquire the traffic congestion information.
Other platforms also aim at the real-time gasoline price (e.g., GasBuddy \cite{GasBuddy} and FuelMap \cite{FuelMap}) and the parking space (e.g., Pavemint \cite{Pavemint} and SpotHero \cite{SpotHero}), to name just a few.
In practice, different platforms (e.g., GasBuddy and FuelMap) could be interested in the same group of PoIs (e.g., gas stations), which forms an interactional status acquisition system.
The system performance primarily depends on \textit{social welfare} and \textit{platform freshness} \cite{abd2019role}, which are the main focus of this paper.
\begin{itemize}
	\item The social welfare is the aggregate payoff of all the PoIs and platforms, which represents the efficiency of the status acquisition system.
	
	\item The platform freshness can be measured by a metric termed \textit{age of information}.
		The age of a platform's content is the time elapses since the generation of the PoI status used in the most recent update of this platform.
\end{itemize}

In general, we want to maintain an \textit{efficient} status acquisition system and keep the platforms  \textit{fresh}.
However, there are many obstacles to achieving this goal.
First, there is an unavoidable trade-off between efficiency and freshness, since a more efficient system operation rule does not necessarily reduce the platform age.
Second, the platforms and PoIs are all self-interested, thus their objectives or preferences may be different or even conflicting.
Specifically, a platform benefits from the status uploaded by the PoIs, while a PoI may hesitate to frequently share its status due to the privacy leakage and other monetary loss.
The self-interested features prevent them from reaching an agreement on how frequent a PoI ought to upload its status to a platform, let alone the efficiency and the freshness.
This motivates us to study the first key question:
\begin{question}\label{Question: 1}
	How to help the self-interested PoIs and platforms reach an efficient and fresh agreement on status acquisition?
\end{question}

\begin{figure}
\setlength{\abovecaptionskip}{5pt}
\setlength{\belowcaptionskip}{0pt}
\centering
\includegraphics[width=0.9\linewidth]{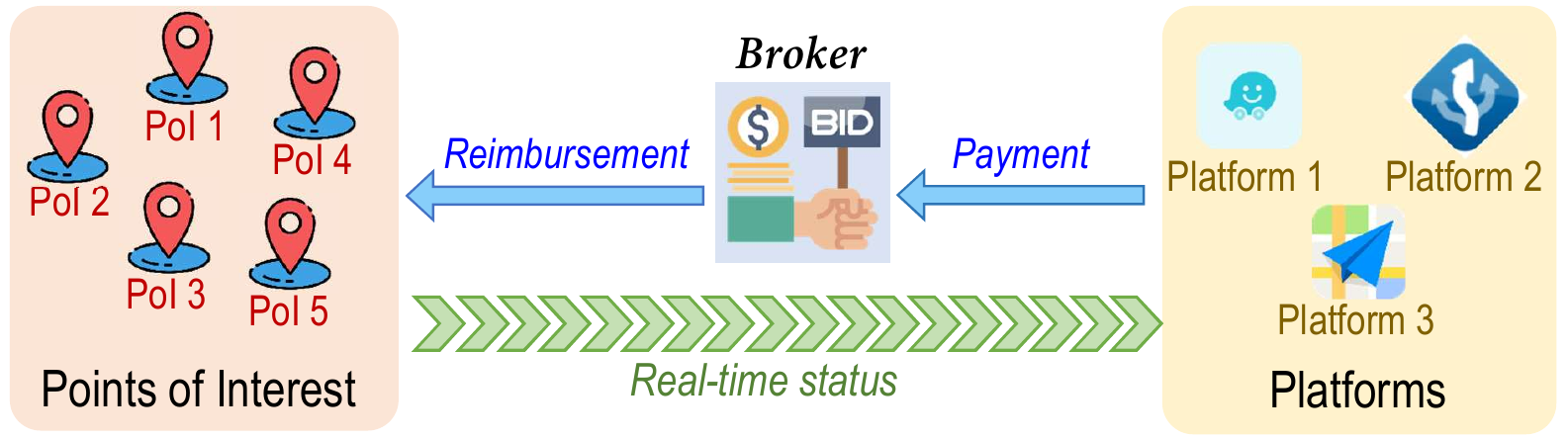}
\caption{An illustrative status acquisition market.}
\label{fig: example}
\end{figure}

One feasible way of resolving {Question \ref{Question: 1}} is to design a market mechanism, such that a broker (e.g., the government) manipulates the marketplace as shown in Fig. \ref{fig: example}.
However, it is challenging to devise a suitable mechanism, since the status acquisition system naturally involves \textit{time-varying information asymmetry}.
Specifically, the platform utility and PoI cost depend on their private information, which can be random and time-varying.
Moreover, updating platform contents usually requires necessary data analytics based on the received PoI status, thus the platform age depends on the updating latency that the broker cannot observe.
This leads to our second key question in this paper:
\begin{question}\label{Question: 2}
	How to elicit the time-varying private information of PoIs and platforms given the efficiency and freshness goal? 
\end{question}

The challenge of Question \ref{Question: 2} is that the private information is ineluctably intertwined with the trade-off between the efficiency and freshness over the multi-period market operation.
Hence one cannot address the instantaneous private information without considering its future impact.
In this paper, we devise a \textit{long-term decomposition (LtD) mechanism}, which assists the broker to coordinate the status acquisition.
We believe that our results in this paper can help promote the efficient and fresh status acquisition in the future.

\subsection{Main Results and Key Contributions}
This paper studies the multi-period operation of a status acquisition system, where each platform desires to obtain the fresh status of PoIs and updates its platform contents.
A market broker operates the marketplace and coordinates the interactions between the self-interested PoIs and platforms.
At the beginning of each period, the broker will help determine how frequently a PoI should upload its real-time status to a platform.
During this period, each platform will update its contents after analyzing the received PoI status.
We aim to design a market mechanism, which maximizes the social welfare and ensures the platform freshness constraints.

The main results and key contributions of this paper are summarized as follows:

\begin{itemize}
	\item \textit{A Novel Problem Formulation:}
	We study the real-time operation of a status acquisition system, where the platforms and PoIs are associated with time-varying private utility and cost, respectively.
	The platform age depends on its updating latency, which is also unknown to the market broker.
	As far as we know, we are the first to study the market operation with time-varying private social welfare and freshness constraints.
	
	\item \textit{Mechanism Design:}
	We devise a long-term decomposition (LtD) mechanism to resolve the time-varying information asymmetry.
	Specifically, we construct a virtual social welfare that only depends on the current private information.
	We then decompose the per-period operation into multiple distributed bidding problems for the PoIs and platforms.
	To the best our knowledge, we design the first market mechanism addressing the time-varying information asymmetry.

	\item \textit{Theoretical Performance:}
	We show that the LtD mechanism retains the same social performance compared to the symmetric information case, and asymptotically ensures the platform freshness conditions.
	Moreover, the broker does not need to inject or take money, thus maintains a balanced budget.
	These properties enable the non-profit broker to run the LtD mechanism, and attract more PoIs and platforms to join the market.

	\item \textit{Extensive Evaluation:}
	We evaluate the LtD mechanism based on real-world electricity price data.
	The numerical results show that when platforms pay more attention to the payoff maximization, each PoI can still obtains a non-negative payoff in the long-term.
\end{itemize}

\subsection{Related Works}
Age of information is a new metric to characterize the information freshness. 
The early studies analyze the average age under different queuing disciplines (e.g., \cite{Sun2017Update,huang2015optimizing,yates2018age}) and age minimization in wireless network (e.g., \cite{kadota2019scheduling,Igor2019Minimizing,Bedewy2019Age}).
This paper is mostly related to the economic age management in online content platforms, thus we focus on this stream of studies next.

The age-based platform operation usually involves the interaction between platforms and PoIs.
Zhang \textit{et al} in \cite{meng2019wiopt} focus on AoI pricing problem, and compare time-dependent pricing and quantity-based pricing.
Wang \textit{et al} \cite{Duan2019ISIT} consider a dynamic pricing problem, where the platform offers age-dependent reward and encourages PoIs to upload their status at different rates.
Li \textit{et al} in \cite{li2019can} design a linear age-based reward and characterize the system efficiency in terms of price of anarchy.
Hao \textit{et al} \cite{hao2020regulating} further take into account multiple platforms that competitively sample the PoI status.
They design a trigger mechanism to ensure the social optimum under the platform cooperation.

This paper differs from \cite{meng2019wiopt,Duan2019ISIT,li2019can,hao2020regulating} in two aspects.
First, we consider a general status acquisition market with multiple self-interested PoIs and platforms.
Second, we take into account time-varying private information for each PoI and platform.
The two aspects above substantially increase the challenge of achieving an efficient and fresh market.

The remainder of the paper is as follows.
Section \ref{Section: System Model} introduces the system model and the problem formulation.
Section \ref{Section: Symmetric Information as Benchmark} derives the benchmark solution under symmetric information scenario.
Section \ref{Section: Asymmetric Information} introduces our proposed LtD mechanism and Section \ref{Section: Performance} presents its theoretical performance.
Section \ref{Section: Numerical results} provides the numerical results.
We conclude this paper in Section \ref{Section: Conclusion}.

\section{System Model}
\label{Section: System Model}
We consider a status acquisition system operated in a set $\mathcal{T}=\{1,2,...,T\}$ of periods.
Each period $t\in\mathcal{T}$ has the same duration $\Delta$ (e.g., one day), and we use $\tau\in[0,T\Delta]$ to denote the continuous time. 
The system consists of a set $\mathcal{N}=\{1,2,...,N\}$ of platforms and a set $\mathcal{I}=\{1,2,...,I\}$ of Points of Interest (PoIs).
\begin{itemize}[leftmargin=13pt]
	\item Each PoI $i\in\mathcal{I}$ is associated with some time-varying status (e.g., congestion or parking space).
	We let $\status_{i}(\tau)$ denote the instantaneous status of PoI $i$ at time $\tau\in[0,T\Delta]$.
	
	\item Each platform $n\in\mathcal{N}$ wants to update its platform contents based on the real-time PoI status.
	We let $\Status_{n,i}(\tau)$ denote platform $n$'s content related to PoI $i$ at time $\tau$.
\end{itemize}

Next we introduce the status acquisition process in Section \ref{Subsection: Status Acquisition Interactions}, and then characterize the platform utility and PoI cost in Section \ref{Subsection: SP Utility PoI Cost}.
We introduce the freshness condition and problem formulation in Section \ref{Subsection: Market Freshness} and Section \ref{Subsection: Problem Formulation}, respectively.

\subsection{Real-Time Status Acquisition}
\label{Subsection: Status Acquisition Interactions}
The real-time status acquisition primarily consists of two phases, i.e., \textit{status uploading} and \textit{platform updating}.

\subsubsection{\textbf{Status Uploading}}
Status uploading specifies how frequently each PoI uploads its real-time status to each platform.
Specifically, we follow the previous studies (e.g., \cite{kaul2012realtime,sun2017up,narasimha2020mean}), and consider that each PoI $i$ maintains a Poisson clock with the normalized rate and can acquire its instantaneous status whenever the clock ticks.
Each PoI $i$ can flexibly decide whether to upload the acquired status to each platform.
In practice, it is costly for a PoI to make the uploading decision repetitively.
Hence we consider a probabilistic uploading scheme, namely, each PoI $i$ uploads its acquired status to platform $n$ with the probability $x^{t}_{i, n}\in[0,1]$ in period $t$.
We refer to $x^{t}_{i, n}$ as the uploading rate from PoI $i$ to platform $n$.
The uploading rate of the entire system in period $t$ is
\begin{equation}
\bm{x}^{t}
\triangleq
\left(x^{t}_{i, n}:\forall i\in\mathcal{I},n\in\mathcal{N}\right),
\end{equation}
which is chosen from the set ${\mathcal{X}}\triangleq[0,1]^{I\times N}$.
We will abuse notation a little bit, and let $\bm{x}^{t}_{n}=(x^{t}_{i,n}:\forall i\in\mathcal{I})$ and  $\bm{x}^{t}_{i}=(x^{t}_{i,n}:\forall n\in\mathcal{N})$ denote the input uploading rate to platform $n$ and the output uploading rate from PoI $i$, respectively.

\subsubsection{\textbf{Platform Updating}}
\label{Subsubsection: Platform Updating}
The platform updates its contents based on the received PoI status, which involves data analytics and inevitably incurs updating latency \cite{alabbasi2020joint}.
The updating latency primarily depends on the computing capability of the platform and the computing requirement of the PoI status.
\begin{itemize}
	\item The computing capability of a platform can be roughly measured based on its CPU frequency (i.e., the number of CPU cycles per second) \cite{chen2015efficient}.
	Accordingly, we let $\proc_{n}$ denote the computing capability of platform $n$.
	
	\item The computing requirement is the required number of CPU cycles to analyze the PoI status.
	Based on the empirical studies (e.g., \cite{lee2018speeding,reisizadeh2019coded,liang2014tofec}), we model the computing requirement as an exponentially distributed random variable with the normalized mean. 
	Our analysis is not limited to the specific distribution, which is elaborated in Section \ref{Subsection: Market Freshness}.
\end{itemize}

The computing capability and the computing requirement jointly determines the platform updating latency, which eventually affects the platform age.
We will introduce the platform age in Section \ref{Subsection: Market Freshness}.
Before that, we first characterize the platform utility and PoI cost in what following.

\subsection{Platform Utility \& PoI Cost}
\label{Subsection: SP Utility PoI Cost}
We first model the utility of each platform $n\in\mathcal{N}$ and the cost of each PoI $i\in\mathcal{I}$.
We then derive the social welfare.

\subsubsection{\textbf{Platform Utility}}
The platform gains utility from the status uploaded by PoIs.
The utility of each platform $n$ primarily depends on two aspects as follows:
\begin{itemize}
	\item First, the platform utility represents the revenue (from advertisements) or the cost reduction (compared to the case where the platform acquires the PoI status itself).
	Therefore, the utility of platform $n$ is positively related to the input uploading rate $\bm{x}^{t}_{n}
	=
	(x^{t}_{i, n}:\forall i\in\mathcal{I})$.

	\item Second, the platform utility also depends on many other factors (e.g., ad price or human resource investment), which are random and time-varying.
	In general, we let $\rdmSP_{n}$ denote the random factor that could potentially affect the utility of platform $n$.
	Mathematically, $\rdmSP_{n}$ is a random variable on the support $\Sigma_{n}$ for platform $n$.
\end{itemize}

Based on the above discussion, we model the utility of platform $n$ in period $t$ as follows:
\begin{equation}
U_{n}\left(\bm{x}^{t}_{n};\rdmSP^{t}_{n}\right)
\triangleq
\sum_{i\in\mathcal{I}}U_{n,i}\left(x^{t}_{i, n}; \rdmSP^{t}_{n}\right),
\end{equation}
where $\rdmSP^{t}_{n}\in\Sigma_{n}$ is the realization of the random factor in period $t$.
To capture the diminishing marginal return, we assume that $U_{n,i}(x;\rdmSP^{t}_{n})$ is concave and increasing in $x$, and satisfies  $U_{n,i}(0;\rdmSP^{t}_{n})=0$ for any realization $\rdmSP^{t}_{n}\in\Sigma_{n}$.


\subsubsection{\textbf{PoI Cost}}
The PoI $i$ incurs cost from uploading its status to the platforms.
We model the PoI cost as follows.
\begin{itemize}
	\item The PoI cost usually consists of monetary cost (e.g., the energy expenditure) and non-monetary cost (e.g., the privacy loss).
	Both of them are positively related to the output uploading rate $\bm{x}^{t}_{i}
	=
	( x^{t}_{i, n}:\forall n\in\mathcal{N} )$ of PoI $i$.

	\item The PoI cost also depends on some other time-varying factors, such as the electricity price or the privacy sensitivity.
	In general, we let $\rdmPoI_{i}$ denote the random factor that may affect the cost of PoI $i$.
	Mathematically, $\rdmPoI_{i}$ is a random variable on the support $\Omega_{i}$.
\end{itemize}

Therefore, we characterize the cost of PoI $i$ in period $t$ as
\begin{equation}
C_{i}(\bm{x}^{t}_{i};\rdmPoI^{t}_{i})
\triangleq
\sum_{n\in\mathcal{N}}C_{i,n}(x^{t}_{i, n};\rdmPoI^{t}_{i}),
\end{equation}
where $\rdmPoI^{t}_{i}\in\Omega_{i}$ is the realization of the random factor in period $t$.
We suppose that the cost function  $C_{i,n}(x;\rdmPoI^{t}_{i})$ is convex and increasing in $x$, and satisfies $C_{i,n}(0;\rdmPoI^{t}_{i})=0$.

Note that the random factors $\rdmSP^{t}_{n}$ and $\rdmPoI^{t}_{i}$ are the private information of platform $n$ and PoI $i$, respectively.
The corresponding utility and cost functions are only known to the platform and the PoI, respectively.
For notation simplicity, we often suppress the dependency on the random factors, and use $U^{t}_{n}(\bm{x})$ and $C^{t}_{i}(\bm{x})$ to denote the private utility and cost functions of platform $n$ and PoI $i$ in period $t$, respectively.

\subsubsection{\textbf{Social Welfare}}
The system social welfare is the difference between the total utility of platforms and the total cost of PoIs.
Specifically, the social welfare in period $t$ is given by
\begin{equation}\label{Equ: SW period}
\begin{aligned}
S(\bm{x}^{t};\bm{\rdmSP}^{t},\bm{\rdmPoI}^{t})
\triangleq
\sum_{n\in\mathcal{N}}\sum_{i\in\mathcal{I}}
U_{n,i}^{t}(x_{i, n}^{t})
-
C_{i,n}^{t}(x_{i,n}^{t}),
\end{aligned}
\end{equation}
which depends on $\bm{\rdmSP}^{t}=(\rdmSP^{t}_{n}:\forall n\in\mathcal{N})$ and $\bm{\rdmPoI}^{t}=(\rdmPoI^{t}_{i}:\forall i\in\mathcal{I})$ in period $t$.
Similarly, we sometimes use $S^{t}(\bm{x}^{t})$ to denote the social welfare in period $t$.

\subsection{Market Freshness}
\label{Subsection: Market Freshness}
The platform desires to keep its platform contents fresh, which is measured by the corresponding age.
Next we define age formally, and then formulate the freshness condition.

\subsubsection{\textbf{Definition of Age}}
The age of the platform content is the time that elapsed since the uploading time of the PoI status used in the most recent updating of this content.
Mathematically, the age of the platform content $\Status_{n,i}(\tau)$ at time $\tau\in[0,T\Delta]$ is 
\begin{equation}\label{Equ: age definition}
a_{n,i}(\tau)
\triangleq
\tau - J_{n,i}(\tau),
\end{equation}
where $J_{n,i}(\tau)$ represents the uploading time of the PoI status used in the most recent updating up to time $\tau$.

\begin{figure}
	\setlength{\abovecaptionskip}{5pt}
	\setlength{\belowcaptionskip}{0pt}
	\centering
	\includegraphics[width=0.7\linewidth]{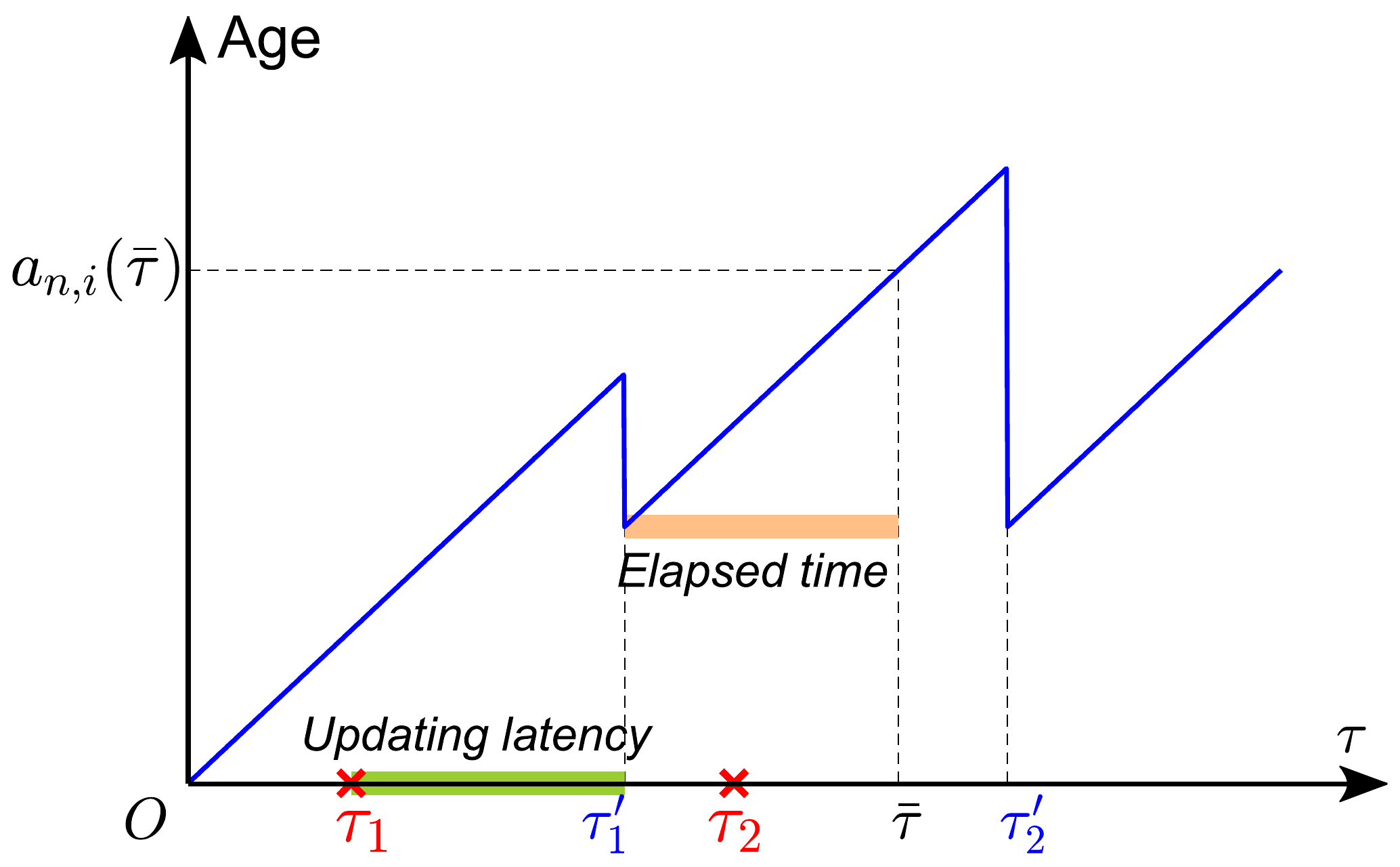}
	\caption{A numerical example of $a_{n,i}(\tau)$}
	\label{fig: Age_illustration}
\end{figure}

We elaborate the definition (\ref{Equ: age definition}) based on a numerical example shown in Fig. \ref{fig: Age_illustration}.
Specifically, the blue curve represents $a_{n,i}(\tau)$, i.e., the age of platform content $\Status_{n,i}(\tau)$.
There are two \textit{status uploading} events (from PoI $i$ to platform $n$) at time $\tau_{1}$ and $\tau_{2}$.
The platform $n$ updates the content (related to PoI $i$) twice at time $\tau'_{1}$ and $\tau'_{2}$.
Overall, the age increases linearly and drops vertically whenever an updating event happens.
Note that we have $J_{n,i}(\bar{\tau})=\tau_{1}$ for any $\bar{\tau}\in[\tau_{1}',\tau_{2}')$.
The age $a_{n,i}(\bar{\tau})=\bar{\tau}-\tau_{1}$ is essentially the sum of the updating latency $\tau'_{1}-\tau_{1}$ (i.e., the green line segment) and the elapsed time $\bar{\tau}-\tau'_{1}$ (i.e., the orange line segment). 
Moreover, the updating latency is a random variable and jointly depends on the platform computing capability and the computing requirement of analyzing the PoI status according to Section \ref{Subsubsection: Platform Updating}.

\subsubsection{\textbf{Platform Age}}
Based on the above discussion, we define the age of platform $n$ as the average age over the platform contents $\{\Status_{n,i}(\tau):\forall i\in\mathcal{I}\}$.
Mathematically, at time $\tau$, the age of platform $n$ is given by 
\begin{equation}
a_{n}(\tau)
\triangleq
\frac{1}{|\mathcal{I}|}\sum_{i\in\mathcal{I}} a_{n,i}(\tau) .
\end{equation}

The platform aims to keep its contents fresh in the long-term.
We quantify the long-term freshness based on the time-average age over the time horizon $[0,T\Delta]$, i.e.,
\begin{equation}\label{Equ: T-period time-average}
\frac{1}{T\Delta}
\int_{0}^{T\Delta} 
a_{n}(\tau) {\rm d}\tau
=\frac{1}{T}\sum_{t=1}^{T}\frac{1}{\Delta}\int_{(t-1)\Delta}^{t\Delta}a_{n}(\tau){\rm d}\tau.
\end{equation}

Note that (\ref{Equ: T-period time-average}) is the \textit{empirical} time-average age of a sample path in the stochastic status acquisition system, which depends on the uploading rates $\{\bm{x}^{t}:\forall t\in\mathcal{T}\}$.
Next we introduce the freshness conditions at the stationary state.

\subsubsection{\textbf{Freshness Condition}}
The status acquisition process in Section \ref{Subsection: Status Acquisition Interactions} implies that each platform $n$ can be viewed as an M/M/1 queuing system with the serving rate $\proc_{n}$ and the $I$-source input uploading rate $\bm{x}^{t}_{n}=(x^{t}_{i,n}:\forall i\in\mathcal{I})$.
Based on the previous study in \cite{yates2018age}, the \textit{stationary} time-average age of platform $n$ is given by
\begin{equation}\label{Equ: AoI}
\begin{aligned}
\AoI_{n}(\bm{x}_{n}^{t})
\triangleq\textstyle \frac{1}{|\mathcal{I}|}\sum\limits_{i\in\mathcal{I}}\frac{1}{\proc_{n}}
\textstyle\Big[
\frac{1}{\rho_{n,i}}
+\frac{1}{1-\rho_{n,-i}}
\textstyle+\frac{\rho_{n,i}^2(1-\rho_{n,i}\rho_{n,-i})}{(1-\rho_{n,i})(1-\rho_{n,-i})^3}
\Big],
\end{aligned}
\end{equation}
where $\rho_{n,i}\triangleq x^{t}_{i, n}/\proc_{n}$ and $\rho_{n,-i}\triangleq\sum_{j\ne i}\rho_{n,j}$.

Note that when the period duration $\Delta$ is large compared to the platform updating latency, the empirical time-average average age in (\ref{Equ: T-period time-average}) is close to the stationary time-average age.
That is, we have the following convergence results
\begin{equation}
\begin{aligned}
	\lim\limits_{\Delta\rightarrow\infty}\frac{1}{T\Delta}\int_{0}^{T\Delta}a_{n}(\tau)
	=\frac{1}{T}
	\sum_{t=1}^{T}
	\AoI_{n}(\bm{x}^{t}_{n}),\quad\forall n\in\mathcal{N}.
\end{aligned}
\end{equation}
We let $\AoIthr_{n}$ denote the freshness threshold of platform $n$, and formulate the following \textit{freshness conditions}:
\begin{equation}\label{Equ: Freshness}
\frac{1}{T}
\sum_{t=1}^{T}
\AoI_{n}(\bm{x}^{t}_{ n})
\le\AoIthr_{n},\ \forall n\in\mathcal{N}.
\end{equation}

Recall that $\AoI_{n}(\cdot)$ in (\ref{Equ: AoI}) is derived based on the exponentially distributed computing requirement.
It is still an open problem to characterize the stationary average age of a queuing system with the general distribution \cite{yates2020age}.
Nevertheless, our later theoretical results only rely on the convexity of $A_{n}(\cdot)$, thus can be potentially extended to the general case.

\subsection{Market Operation Problem}
\label{Subsection: Problem Formulation}
In a status acquisition system, it is difficult for the self-interested PoIs and platforms to reach an agreement on the uploading rate $\bm{x}^{t}$.
Hence it is necessary for a market broker to help coordinate their interactions.
To maintain an efficient and fresh market, the broker needs to solve the following market operation problem.
\begin{problem}[Broker's Market Operation Problem\vspace{-3pt}]\label{Problem: SWM}
	\begin{equation}
		\begin{aligned}
		\max&\   \sum_{t=1}^{T} S^{t}(\bm{x}^{t}) \\
		\textit{s.t. }&\  (\ref{Equ: Freshness}),\\
		\textit{ var. }&\  
		(\bm{x}^{t}\in\mathcal{X}:\ \forall t).
		\end{aligned}
	\end{equation}
\end{problem}

The major challenge for a broker to solve Problem \ref{Problem: SWM} is the time-varying information asymmetry:
First, the social welfare $S^{t}(\cdot)$ is unknown to the broker, as it depends on the private information of PoIs and platforms.
Second, the broker does not know the computing capability of each platform.
In this case, although the broker can observe the platform age at the end of a period, the broker cannot make the decisions based on the age function  $A_{n}(\cdot)$. 
Third, the uploading rates in different periods couple with each other due to the long-term freshness conditions (\ref{Equ: Freshness}).
Hence the broker cannot resolve the per-period information asymmetry independently.

To resolve the challenge above, we devise a \textit{long-term decomposition (LtD) mechanism} in Section \ref{Section: Asymmetric Information}.
Before introducing it, we briefly present the solution of Problem \ref{Problem: SWM} under symmetric information in Section \ref{Section: Symmetric Information as Benchmark}, which is the benchmark of our proposed LtD mechanism.

\section{Symmetric Information Benchmark}
\label{Section: Symmetric Information as Benchmark}
This section focuses on the symmetric information scenario.
That is, the broker knows the platform age function $A_{n}(\cdot)$ and also has the knowledge on the random factors $(\bm{\rdmSP},\bm{\rdmPoI})$.
The knowledge on the random factors has two levels:
\begin{itemize}
	\item \textit{Stochastic Knowledge:}
		The broker can observe the realization $(\bm{\rdmSP}^{t},\bm{\rdmPoI}^{t})$ at the beginning of period $t$, and also possesses the distribution of the random factors $(\bm{\rdmSP},\bm{\rdmPoI})\in\Sigma_{\mathcal{N}}\times\Omega_{\mathcal{I}}$, where $\Sigma_{\mathcal{N}}\triangleq\Sigma_{1}\times\Sigma_{2}\times...\times\Sigma_{N}$ and $\Omega_{\mathcal{I}}\triangleq\Omega_{1}\times\Omega_{2}\times...\times\Omega_{I}$.
	
	\item \textit{Online Observation:}
	The broker can observe the realization $(\bm{\rdmSP}^{t},\bm{\rdmPoI}^{t})$ at the beginning of each period $t$.
	
\end{itemize}

The case \textit{stochastic knowledge} enables the broker to achieve a better performance than the case \textit{online observation}.
Next we introduce how to solve Problem \ref{Problem: SWM} in the two cases.

\subsection{Symmetric Information with Stochastic Knowledge}
When the broker possesses the distribution of the random factors $(\bm{\rdmSP},\bm{\rdmPoI})$, it can determine the uploading rate $\bm{x}$ according to some randomized policy and optimize the policy based on the distribution.
We defined a randomized policy $\bm{\beta}$ as follows:
\begin{equation}
	\bm{\beta}\triangleq
	\big\{
	\beta_{\bm{x}}(\bm{\rdmSP},\bm{\rdmPoI})\in[0,1]:
	\bm{x}\in\mathcal{X},\bm{\rdmSP}\in\Sigma_{\mathcal{N}},\bm{\rdmPoI}\in\Omega_{\mathcal{I}}
	\big\},
\end{equation}
which represents that the broker adopts the uploading rate $\bm{x}\in\mathcal{X}$ with the probability $\beta_{\bm{x}}(\bm{\rdmSP},\bm{\rdmPoI})$ after observing the realization $(\bm{\rdmSP},\bm{\rdmPoI})$.
Accordingly, we let $\bm{\bar{x}}^{\bm{\beta}}\in\mathcal{X}$ denote the random variable following the distribution $\bm{\beta}$.
The broker can obtain the optimal randomized policy $\bm{\beta^*}$ by solving
\begin{equation}\label{Equ: randomized policy}
\begin{aligned}
	\bm{\beta^*}\triangleq\arg
	\max\limits_{\bm{\beta}}&\quad \mathbb{E}\left[S(\bm{\bar{x}}^{\bm{\beta}};\bm{\rdmSP},\bm{\rdmPoI})\right]\\
	\textit{s.t.}&\quad \mathbb{E} \left[\AoI_{n}\big(\bm{\bar{x}}^{\bm{\beta}}\big)\right] \le \AoIthr_{n},\ \forall n\in\mathcal{N},
\end{aligned}
\end{equation}
where the expectation is taken over $\{\bm{\bar{x}}^{\bm{\beta}},\bm{\rdmSP},\bm{\rdmPoI}\}$.
We let $S^*$ denote the expected social welfare achieved by $\bm{\beta^*}$.
Mathematically, we have $S^*=\sum_{n=1}^{N}S^*_{n}$, where $S^*_{n}$ is the social welfare component related to platform $n$, i.e.,
\begin{equation}\label{Equ: SWn}
S^{*}_{n}
\triangleq
\textstyle
\sum\limits_{i\in\mathcal{I}}\mathbb{E}
\left[U_{n,i}\big(\bar{x}_{i, n}^{\bm{\beta^*}};\bm{\rdmSP}\big)
-
C_{i,n}\big(\bar{x}_{i,n}^{\bm{\beta^*}};\bm{\rdmPoI}\big)
\right]
.
\end{equation}

In Section \ref{Section: Performance}, we will elaborate how $S^*_{n}$ affects the payoff of platform $n$ under our proposed LtD mechanism.
Before that, we first introduce the case online observation.

\subsection{Symmetric Information with Online Observation}
When the broker can only observe the current realization $(\bm{\rdmSP}^{t},\bm{\rdmPoI}^{t})$, it can leverage the Lyapunove optimization framework to determine the uploading rate based on the self-defined virtual queues \cite{neely2010stochastic}.
Based on the freshness constraints (\ref{Equ: Freshness}), the broker can define the following virtual queues
\begin{equation}\label{Equ: Virtual Queue}
	\begin{aligned}
		Q_{n}^{t+1}
		=
		\left[
		Q_{n}^{t}+\AoI_{n}\left(\bm{x}^{t}_{n}\right) - \AoIthr_{n}
		\right]^+,\ \forall n\in\mathcal{N},
	\end{aligned}
\end{equation}
where $[\cdot]^+\triangleq\max(\cdot,0)$.
Accordingly, the broker obtains the following \textit{virtual social welfare} in period $t$:
\begin{equation}\label{Equ: virtual SW}
\tilde{S}^{t}(\bm{x};\V)
\triangleq
S^{t}(\bm{x}) - \frac{1}{\V}\sum_{n\in\mathcal{N}}Q^{t}_{n}\AoI_{n}(\bm{x}_{n}),
\end{equation}
where $\V\in(0,+\infty)$ is determined by the broker.

The Lyapunov drift theorem (Theorem 4.1 in \cite{neely2010stochastic}) indicates that if the broker determines the uploading rate according to
\begin{equation}\label{Equ: maximize virtual social welfare}
\begin{aligned}
\bm{\tilde{x}}^{t}
\triangleq
\arg
\max\limits_{\bm{x}\in\mathcal{X}}\ \tilde{S}^{t}(\bm{x};\V) ,
\quad\forall t\in\mathcal{T},
\end{aligned}
\end{equation}
then the solution $\{\bm{\tilde{x}}^{t}:\forall t\in\mathcal{T}\}$ achieves a desired performance summarized in Theorem \ref{Theorem: DPP}.
The proof is rather standard and follows the rationale of Chapter 4 in \cite{neely2010stochastic}.
\begin{theorem}\label{Theorem: DPP}
	For any $\V\in(0,+\infty)$, the solution $\{\bm{\tilde{x}}^{t}:\forall t\in\mathcal{T}\}$ achieves the following performance
	\begin{equation}
	\begin{aligned}
	&\lim\limits_{T\rightarrow\infty}\frac{1}{T}\textstyle
	\sum\limits_{t=1}^{T}\mathbb{E}\left[
	\AoI_{n}(\bm{\tilde{x}}^{t})
	\right]
	\le\AoIthr_{n},\ \forall n\in\mathcal{N},\\
	&\lim\limits_{T\rightarrow\infty}\frac{1}{T}\textstyle
	\sum\limits_{t=1}^{T}\mathbb{E}\left[
	S^{t}(\bm{\tilde{x}}^{t})
	\right]
	\ge	
	S^* - O\left(\frac{1}{\V}\right).
	\end{aligned}
	\end{equation}
	where $S^*$ is given in (\ref{Equ: randomized policy}).
\end{theorem}

Theorem \ref{Theorem: DPP} shows that the uploading rates that maximize the virtual social welfare can asymptotically perform as good as the optimal randomized policy $\bm{\beta^*}$.
Nevertheless, (\ref{Equ: maximize virtual social welfare}) requires the knowledge on the realization $(\bm{\rdmSP}^{t},\bm{\rdmPoI}^{t})$.
In Section \ref{Section: Asymmetric Information}, we view $\{\bm{\tilde{x}}^{t}:\forall t\in\mathcal{T}\}$ as the \textit{\textbf{desired solution}} and introduce how to implement it without knowing the random factors.

\section{Asymmetric Information Problem}
\label{Section: Asymmetric Information}
This section will propose a \textit{long-term decomposition (LtD) mechanism} to solve Problem \ref{Problem: SWM} under asymmetric information.
The key idea is to leverage the self-interested feature of PoIs and platforms, and decompose the original market operation problem such that the PoIs and platforms can make the decisions in a distributed manner under a broker's mild coordinations.
We will overview the general rationale of the LtD mechanism in Section \ref{Subsection: Decomposition}. 
We then proceed the detailed design in Sections \ref{Subsection: Bidding Problems}$\sim$\ref{Subsection: Iterative Updating}.

\subsection{Rationale of LtD Mechanism}
\label{Subsection: Decomposition}
The LtD mechanism builds upon a consistent decoupling and an auction scheme, which are introduced in the following.

\subsubsection{\textbf{Consistent Decoupling}}
The virtual social welfare (\ref{Equ: virtual SW}) implies that the uploading rate is jointly related to the platform utility and PoI cost, as well as the platform age.
That is, the uploading rate couples the private information of the platforms and PoIs.
To decouple it, we introduce a set of auxiliary variables $\bm{y}\in\mathcal{X}$ and the following \textit{consistency constraints}:
\begin{equation}\label{Equ: feasibility}
\begin{aligned}
x_{i,n} &= y_{i,n},\ \forall n\in\mathcal{N},i\in\mathcal{I}.
\end{aligned}
\end{equation}

The auxiliary variables $\bm{y}$ enable us to define the following \textit{decoupled virtual social welfare}:
\begin{equation}\label{Equ: decomposed virtual social welfare}
\begin{aligned}
\tilde{S}^{t}_{\textit{D}}(\bm{x},\bm{y};V)
\triangleq
\sum\limits_{n\in\mathcal{N}}\left[U_{n}^{t}(\bm{x}_{n}) -\frac{Q^{t}_{n}\AoI_{n}(\bm{x}_{n})}{\V}\right]
-\sum\limits_{i\in\mathcal{I}}C_{i}^{t}(\bm{y}_{i}).
\end{aligned}
\end{equation}
Accordingly, the virtual social welfare maximization problem in (\ref{Equ: maximize virtual social welfare}) can be equivalently transformed into
\begin{equation}\label{Equ: De-v-SW Maximziation}
\begin{aligned}
(\bm{\tilde{x}}^{t},\bm{\tilde{y}}^{t})
\triangleq
\arg
\max\limits_{\bm{x},\bm{y}\in\mathcal{X}}&\quad  \tilde{S}^{t}_{\textit{D}}(\bm{x},\bm{y};\V)\\
\textit{s.t. }	&\quad  (\ref{Equ: feasibility}).
\end{aligned}
\end{equation}

Note that (\ref{Equ: De-v-SW Maximziation}) is a convex optimization problem, thus one can obtain the optimal solution based on
Karush-Kuhn-Tucker (KKT) conditions.
We let $\bm{\lambda}=(\lambda_{i,n}:\forall n\in\mathcal{N},i\in\mathcal{I})$ denote the dual variables associated with the consistency constraints (\ref{Equ: feasibility}).
In particular, $\lambda_{i,n}$ is also known as the \textit{``consistency price''} in the related studies (e.g., \cite{tan2006distributed}).
We characterize the optimal solution $(\bm{\tilde{x}}^{t},\bm{\tilde{y}}^{t},\bm{\tilde{\lambda}}^{t})$ of (\ref{Equ: De-v-SW Maximziation}) based on the following KKT conditions:
\begin{equation}\label{Equ: KKT original}
	\begin{aligned}
		\frac{\partial U^{t}_{n,i}(\tilde{x}_{i,n}^{t})}{\partial x_{i,n}}
		-\frac{Q^{t}_{n}}{V}\cdot\frac{\partial \AoI_{n}(\bm{\tilde{x}}_{n}^{t})}{\partial x_{i,n}} 
		&= \tilde{\lambda}_{i,n}^{t} ,\quad\forall n,i,\\
		\frac{\partial C^{t}_{i,n}(\tilde{y}^{t}_{i,n})}{\partial y_{i, n}}  
		&= \tilde{\lambda}^{t}_{i,n},\quad\forall n,i,\\
		\tilde{x}^{t}_{i,n} 
		&= \tilde{y}^{t}_{i,n},\quad\forall i,n,
	\end{aligned}
\end{equation}
which depends on the private gradient information of the platform utility, the PoI cost, and the platform age.
Therefore, the broker has to elicit the above private information to derive $(\bm{\tilde{x}}^{t},\bm{\tilde{y}}^{t},\bm{\tilde{\lambda}}^{t})$.
Next we introduce how to achieve this goal.

\subsubsection{\textbf{Auction Scheme}}
To elicit the aforementioned private information, the LtD mechanism decomposes (\ref{Equ: De-v-SW Maximziation}) into multiple distributed bidding problems for the platforms and PoIs through an auction scheme.
Auction \ref{Auction: } summarizes the major procedure with the general \textit{allocation rule} and \textit{payment rule}.
\begin{auction}\label{Auction: }
	In period $t$, the broker announces the allocation function $X^{t}_{i,n}(\cdot)$ and the pricing function $\Pi^{t}_{n}(\cdot)$ for each platform $n\in\mathcal{N}$, as well as the allocation function $Y^{t}_{i,n}(\cdot)$ and the reimbursement function $\Phi^{t}_{i}(\cdot)$ for each PoI $i\in\mathcal{I}$.
	\begin{itemize}[leftmargin=13pt]		
		\item Each platform $n$ and PoI $i$ submit the bid $\bm{\bidSP}^{t}_{n}=(\bidSP^{t}_{i,n}:\forall i\in\mathcal{I})$ and $\bm{\bidPoI}_{i}^{t}=(\bidPoI_{i,n}^{t}:\forall n\in\mathcal{N})$, respectively.
		
		\item The broker determines the uploading rate according to
		\begin{equation}\label{Equ: Acution(x,y)}
			\begin{aligned}
				x^{t}_{i,n} &= X^{t}_{i,n}(\bidSP^{t}_{i,n}),\\
				y^{t}_{i, n} &= Y^{t}_{i, n}(\bidPoI^{t}_{i, n}),
			\end{aligned}
			\quad\forall n\in\mathcal{N},i\in\mathcal{I}.
		\end{equation}
		The broker charges the price $\Pi^{t}_{n}(\bm{\bidSP}^{t}_{n})$ from  platform $n\in\mathcal{N}$ and pays the reimbursement $\Phi^{t}_{i}(\bm{\bidPoI}^{t}_{i})$ to PoI $i\in\mathcal{I}$.
	\end{itemize}
\end{auction}

Auction \ref{Auction: } leads to a bidding problem for each platform and PoI.
Our next task is to design the allocation rule (i.e., $X^{t}_{i,n}(\cdot)$ and $Y^t_{i,n}(\cdot)$) and the payment rule (e.g., $\Pi^{t}_{n}(\cdot)$ and $\Phi^{t}_{i}(\cdot)$), such that the platforms and PoIs will truthfully disclose the gradient information in (\ref{Equ: KKT original}).
We will proceed in three steps:
\begin{itemize}
	\item \textbf{Section \ref{Subsection: Bidding Problems}:} 
	We first formulate the bidding problems of PoIs and platforms in Problem \ref{Problem: PoI} and Problem \ref{Problem: SP_t}, respectively.
	We then investigate their optimal bidding strategies given the general allocation and payment rules.
	
	\item \textbf{Section \ref{Subsection: Allocation Rule}}: 
	We propose the allocation rule $X^{t}_{i,n}(\cdot)$ and $Y^{t}_{i,n}(\cdot)$ by solving the broker's allocation problem, i.e., Problem \ref{Problem: Broker Allocation}.
	We then derive three conditions in Proposition \ref{Proposition: KKT equivalent}, under which the optimal bidding strategies satisfy the KKT conditions (\ref{Equ: KKT original}) given the above allocation rule.
	
	\item \textbf{Section \ref{Subsection: Payment Rule}}: 
	We design the payment rule $\Pi^{t}_{n}(\cdot)$ and $\Phi^{t}_{i}(\cdot)$ guided by the conditions in Proposition \ref{Proposition: KKT equivalent}.
\end{itemize}

\subsection{Bidding Problems \& Strategies}
\label{Subsection: Bidding Problems}
\subsubsection{\textbf{PoI Bidding}}
The payoff of a PoI is the difference between the reimbursement and the cost, thus the self-interested PoI has the following bidding problem:
\begin{tcolorbox}
\begin{problem}[\small Bidding Problem of PoI $i$ in Period $t$\vspace{-3pt}]\label{Problem: PoI}
\begin{equation*}
\begin{aligned}
\bm{\hat{\bidPoI}}^{t}_{i}
\triangleq\arg
	\max&\quad \Phi^{t}_{i}(\bm{\bidPoI}_{i}) - C^{t}_{i}(\bm{y}_{i})  \\
	\textit{ s.t.}&\quad  y_{i, n}=Y^{t}_{i, n}(\bidPoI_{i, n}),\  \forall n\in\mathcal{N},\\
	\textit{ var.}&\quad  \bm{\bidPoI}_{i}=(\bidPoI_{i, n}:\forall n\in\mathcal{N}).
\end{aligned}
\end{equation*}
\end{problem}
\end{tcolorbox}

We let $\bm{\hat{\bidPoI}}_{i}\triangleq(\hat{\bidPoI}^{t}_{i, n}:\forall n\in\mathcal{N})$ and $\bm{\hat{y}}_{i}\triangleq(\hat{y}^{t}_{i, n}:\forall n\in\mathcal{N})$ denote the optimal bid and the corresponding output uploading rate of PoI $i$, respectively.
As we will see later, our proposed allocation and reimbursement rules can ensure that Problem \ref{Problem: PoI} is convex.
Hence we derive the optimal bidding strategy based on the following optimality condition:
\begin{equation}\label{Equ: Optimal Bidding Strategy PoI}
\frac{\partial\Phi^{t}_{i}(\bm{\hat{\bidPoI}}^{t}_{i})}{\partial \bidPoI_{i,n}}
=
\frac{\partial C^{t}_{i,n}(\hat{y}^{t}_{i,n})}{\partial y_{i,n}}
\cdot
\frac{\partial Y^{t}_{i}(\hat{\bidPoI}^{t}_{i,n})}{\partial \bidPoI_{i,n}},
\ \forall i\in\mathcal{I},
\end{equation}
which will be used to design the payment rule in Section \ref{Subsection: Payment Rule}.

\subsubsection{\textbf{Platform Bidding}}
The platform payoff is the difference between the utility and the price (charged by the broker).
Different from PoIs, the platform aims to maximize its payoff and keep its contents fresh.
Hence each platform $n$ has the following long-term bidding problem.
\begin{problem}[Long-term Bidding Problem of Platform $n$\vspace{-3pt}]\label{Problem: SP}
\begin{subequations}
\begin{align}	
\max\limits_{\{\bm{\bidSP}^{t}_{n}\}_{t\in\mathcal{T}}}
&\quad\textstyle \sum\limits_{t=1}^{T}
\left[U^{t}_{n}(\bm{x}^{t}_{n})-\Pi^{t}_{n}(\bm{\bidSP}^{t}_{n})\right]\\
\textit{ s.t. }&\quad  x^{t}_{i,n}=X_{i,n}^{t}(\bidSP^{t}_{i,n}),\ \forall i\in\mathcal{I},\label{Equ: SP Bidding pi} \\
&\quad \frac{1}{T}\textstyle\sum\limits_{t=1}^{T}\AoI_{n}(\bm{x}^{t}_{n})\le\AoIthr_{n}.
\end{align}
\end{subequations}
\end{problem}

To solve Problem \ref{Problem: SP}, the platform can leverage the Lyapunove optimization and define the following virtual queue
\begin{equation}\label{Equ: Virtual Queue q}
q_{n}^{t+1}
=\left[q_{n}^{t}+\AoI_{n}(\bm{x}^{t}_{n}) - \AoIthr_{n}\right]^+,
\end{equation}
where $\bm{x}^{t}_{n}$ is specified according to (\ref{Equ: SP Bidding pi}).
In general, the virtual queues in (\ref{Equ: Virtual Queue q}) and (\ref{Equ: Virtual Queue}) could be different.
As we will see later, our proposed LtD mechanism can ensure $Q_{n}^{t}=q_{n}^{t}$ for any platform $n\in\mathcal{N}$ in each period $t\in\mathcal{T}$.

Based on the Lyapunove method, the platform $n$ can determine its bid in each period $t$ by solving Problem \ref{Problem: SP_t}.
\begin{tcolorbox}
\begin{problem}[\small Bidding Problem of Platform $n$ in Period $t$\vspace{-5pt}]\label{Problem: SP_t}
\begin{equation*}
\begin{aligned}
\bm{\hat{\bidSP}}^{t}_{n}
\triangleq
\arg
\max& \quad  U^{t}_{n}(\bm{x}_{n}) - \Pi^{t}_{n}(\bm{\bidSP}_{n}) -\frac{q^{t}_{n}\AoI_{n}(\bm{x}_{n})}{\V} \\
\textit{s.t.}&\quad x_{i,n}=X^{t}_{i,n}(\bidSP_{i,n}), \ \forall i\in\mathcal{I},\\
\textit{var.}&\quad \bm{\bidSP}_{n}=(\bidSP_{i,n}:\forall i\in\mathcal{I}). 
\end{aligned}
\end{equation*}
\end{problem}
\end{tcolorbox}

We let $\bm{\hat{\bidSP}}^{t}_{n}\triangleq(\hat{\bidSP}^{t}_{i,n}:\forall i\in\mathcal{I})$ and $\bm{\hat{x}}^{t}_{n}\triangleq(\hat{x}^{t}_{i,n}:\forall i\in\mathcal{I})$ denote the optimal bid and the corresponding input uploading rate of platform $n$, respectively.
We derive the optimal bidding strategy based on the following optimality condition:
\begin{equation}\label{Equ: Optimal Bidding Strategy SP}
\frac{\partial \Pi_{n}^{t}(\bm{\hat{\bidSP}}_{n}^{t})}{\partial\bidSP_{i,n}}
=
\frac{\partial X^{t}_{i,n}(\tilde{\bidSP}^{t}_{i,n})}{\partial\bidSP_{i,n}}
\bigg[\frac{\partial U_{n,i}^{t}(\hat{x}_{i,n}^{t})}{\partial x_{i,n}}-\frac{q^{t}_{n}}{\V}\frac{\AoI_{n}(\bm{\hat{x}}^{t}_{n})}{\partial x_{i,n}}\bigg],
\end{equation}
which will be used to design the payment rule in Section \ref{Subsection: Payment Rule}.


\subsection{Allocation Rule}
\label{Subsection: Allocation Rule}
In Auction \ref{Auction: }, the broker determines the uploading rate according to the allocation rule $X^{t}_{i,n}(\cdot)$ and $Y^{t}_{i,n}(\cdot)$.
We follow the previous studies on network utility maximization (e.g., \cite{kelly1998rate,palomar2006tutorial,iosifidis2014double}) and design the allocation rule based on a logarithmic and a quadratic functions.
Specifically, given the platforms' bids $(\bm{\bidSP}^{t}_{n}:\forall n\in\mathcal{N})$ and the PoIs' bids $(\bm{\bidPoI}^{t}_{i}:\forall i\in\mathcal{I})$, the broker determines the uploading rate by solving the following allocation problem.
\begin{tcolorbox}
\begin{problem}[\small Broker's Allocation Problem in Period $t$\vspace{-5pt}]
	\label{Problem: Broker Allocation}
	\begin{equation*}
	\begin{aligned}
	\max\limits_{\bm{x},\bm{y}\in\mathcal{X}}&\quad \sum\limits_{n\in\mathcal{N}}
	\sum\limits_{i\in\mathcal{I}}
	\left[{\bidSP}_{i,n}^{t}\log(x_{i,n}) - \frac{{\bidPoI}^{t}_{i, n}}{2}y_{i, n}^2\right] \\
	\textit{ s.t. }&\quad  (\ref{Equ: feasibility}).
	\end{aligned}
	\end{equation*}
\end{problem}
\end{tcolorbox}

Problem \ref{Problem: Broker Allocation} provides a guideline for us to design the allocation rule, i.e., $X^{t}_{i,n}(\cdot)$ and $Y^{t}_{i,n}(\cdot)$.
To see this, we first express the KKT conditions of Problem \ref{Problem: Broker Allocation} as follows:
\begin{equation}\label{Equ: KKT auction}
\begin{aligned}
{\bidSP^{t}_{i,n}}
= \hat{\lambda}^{t}_{i,n}\hat{x}^{t}_{i,n},\ \  
\bidPoI^{t}_{i, n}\hat{y}^{t}_{i,n} 
= \hat{\lambda}^{t}_{i,n},\ \ 
\hat{x}^{t}_{i,n}
=\hat{y}^{t}_{i,n},\  \forall n,i,
\end{aligned}
\end{equation}
where $(\hat{x}^{t}_{i,n},\hat{y}^{t}_{i,n})$ and $\hat{\lambda}^{t}_{i,n}$ are the optimal solution and dual variables of Problem \ref{Problem: Broker Allocation}, respectively.
The KKT conditions (\ref{Equ: KKT auction}) motivative us to adopt the following \textit{allocation rule}:
\begin{equation}\label{Equ: Allocation functions}
	\begin{aligned}
		X^{t}_{i,n}(\bidSP_{i,n}) 
		\triangleq \frac{\bidSP_{i,n}}{{\lambda}^{t}_{i,n}}
		\quad \text{and}\quad
		Y^{t}_{i, n}(\bidPoI_{i,n}) 
		\triangleq \frac{{\lambda}^{t}_{i,n}}{\bidPoI_{i,n}},\quad\forall n,i,
	\end{aligned}
\end{equation}
which are parameterized by $\bm{\lambda}^{t}=(\lambda^{t}_{i,n}:\forall i\in\mathcal{I},n\in\mathcal{N})$.
Proposition \ref{Proposition: KKT equivalent} elaborates why this allocation rule is good and how to set the parameters $\bm{\lambda}^{t}$.

\begin{proposition}\label{Proposition: KKT equivalent}
	Given the allocation rule in (\ref{Equ: Allocation functions}), we have $(\bm{\hat{x}}^{t},\bm{\hat{y}}^{t})=(\bm{\tilde{x}}^{t},\bm{\tilde{y}}^{t})$, if the following three conditions hold:
	\begin{enumerate}[leftmargin=15pt]		
		\item The optimal bid $\hat{\bidSP}^{t}_{i,n}$ of platform $n$ in Problem \ref{Problem: SP_t} and the uploading rate $\hat{x}^{t}_{i,n}=X^{t}_{i,n}(\hat{\bidSP}^{t}_{i,n})$ satisfy
		\begin{equation}\label{Equ: desired bid SP} 
			\hat{\bidSP}^{t}_{i,n}
			= 
			\hat{x}^{t}_{i,n}
			\left[
			\frac{\partial U^{t}_{n,i}(\hat{x}^{t}_{i,n}) }{\partial x_{i,n}}
			-\frac{Q^{t}_{n}}{V}\frac{\partial \AoI_{n}(\bm{\hat{x}}_{n}^{t})}{\partial x_{i,n}} 
			\right],\ \forall n,i.
		\end{equation}
		
		\item The optimal bid $\hat{\bidPoI}^{t}_{i,n}$ of PoI $i$ in Problem \ref{Problem: PoI} and the uploading rate $\hat{y}^{t}_{i,n}=Y^{t}_{i,n}(\hat{\bidPoI}_{i,n}^{t})$ satisfy
		\begin{equation}\label{Equ: desired bid PoI}
			\hat{\bidPoI}^{t}_{i, n}
			= 
			\frac{1}{\hat{y}^{t}_{i, n}}\cdot \frac{\partial C^{t}_{i,n}(\hat{y}^{t}_{i, n})}{\partial y_{i, n}},
			\ \forall n,i.
		\end{equation}
		
		\item The parameters $\bm{\lambda}^{t}$ in allocation rule (\ref{Equ: Allocation functions}) is given by
		\begin{equation}\label{Equ: desired dual variables}
			{\lambda}^{t}_{i,n}
			=
			\tilde{\lambda}^{t}_{i,n},\ \forall n,i.
		\end{equation}
		
	\end{enumerate}
\end{proposition}

One can prove Proposition \ref{Proposition: KKT equivalent} by showing that (\ref{Equ: desired bid SP})-(\ref{Equ: desired dual variables}) are mathematically equivalent to the KKT conditions (\ref{Equ: KKT original}).
Overall, the broker can implement the \textit{desired solution} $(\bm{\tilde{x}}^{t},\bm{\tilde{y}}^{t})$ in asymmetric information if (\ref{Equ: desired bid SP})-(\ref{Equ: desired dual variables}) hold.
Specifically, (\ref{Equ: desired bid SP}) and (\ref{Equ: desired bid PoI}) are the truthfully bidding requirement, while (\ref{Equ: desired dual variables}) specifies the required consistency price.
We will design the payment rule to ensure the truthfulness in Section \ref{Subsection: Payment Rule}.
We then address the consistency price in Section \ref{Subsection: Iterative Updating}.

\subsection{Payment Rule}
\label{Subsection: Payment Rule}
To ensure the truthfulness requirement, we will carefully design the reimbursement function $\Phi_{i}^{t}(\cdot)$ and the pricing function $\Pi^{t}_{n}(\cdot)$ based on the optimal bidding strategies in (\ref{Equ: Optimal Bidding Strategy PoI}) and (\ref{Equ: Optimal Bidding Strategy SP}), respectively.

\subsubsection{\textbf{Reimbursement Function}} 
We present how to design the reimbursement function $\Phi_{i}^{t}(\cdot)$ for each PoI $i$ in Lemma \ref{Lemma: Reward PoI}.
The proof follows from substituting (\ref{Equ: Designed reward function}) into (\ref{Equ: Optimal Bidding Strategy PoI}).
\begin{lemma}\label{Lemma: Reward PoI}
	Given the allocation rule in (\ref{Equ: Allocation functions}), the truthfully bidding condition (\ref{Equ: desired bid PoI}) holds if the reimbursement function is
	\begin{equation}\label{Equ: Designed reward function}
	\Phi^{t}_{i}(\bm{\bidPoI}_{i})
	\triangleq
	\textstyle
	\sum\limits_{n\in\mathcal{N}}{({\lambda}^{t}_{i,n})^2}\big/{\bidPoI_{i, n}}.
	\end{equation}
\end{lemma}

Note that we can equivalently express the above reimbursement function as $\Phi^{t}_{i}(\bm{\bidPoI}_{i})=\sum_{n\in\mathcal{N}}\lambda_{i,n}Y^{t}_{i,n}(\bidPoI_{i,n})$, which implies that a PoI's reimbursement is proportional to its output uploading rate.
Moreover, we will introduce how the broker sets ${\lambda}^{t}_{i,n}$ in Section \ref{Subsection: Iterative Updating}.

\subsubsection{\textbf{Pricing Function}}
\label{Subsection: SP's Payment Function}
We present how to design the pricing function $\Pi^{t}_{n}(\cdot)$ for each platform $n$ in Lemma \ref{Lemma: Payment SP}.
One can prove this lemma by substituting (\ref{Equ: Designed payment function}) into (\ref{Equ: Optimal Bidding Strategy SP}).
\begin{lemma}\label{Lemma: Payment SP}
	Given the allocation rule in (\ref{Equ: Allocation functions}), the truthfulness condition (\ref{Equ: desired bid SP}) holds if $q^{t}_{n}=Q^{t}_{n}$ and the pricing function is 
	\begin{equation}\label{Equ: Designed payment function}
		\Pi^{t}_{n}
		\left(\bm{\bidSP}_{n}\right)
		\triangleq
		\textstyle
		\sum\limits_{i\in\mathcal{I}}\bidSP_{i,n}.
	\end{equation}
\end{lemma}

We have two-fold elaborations on Lemma \ref{Lemma: Payment SP}.
\begin{itemize}[leftmargin=13pt]	
	\item First, we can equivalently express the above pricing function as $\Pi^{t}_{i}(\bm{\bidSP}_{n})=\sum_{i\in\mathcal{I}}\lambda_{i,n}X^{t}_{i,n}(\bidSP_{i,n})$, which shows that the payment of a platform is proportional to its input uploading rate.
	
	\item Second, Lemma \ref{Lemma: Payment SP} requires the virtual queue backlogs are the same, i.e., $Q^{t}_{n}=q^{t}_{n}$.
	Note that we have $Q^{t}_{n}=q^{t}_{n}$ if $(\bm{\hat{x}}^{m},\bm{\hat{y}}^{m})=(\bm{\tilde{x}}^{m},\bm{\tilde{y}}^{m})$ for any $m<t$.
	As we will see later, this is true under our proposed LtD mechanism.
\end{itemize}

So far, we have introduced the allocation and payment rules, both of which are closely related to the consistency price $\bm{\lambda}^{t}$.
Next we introduce how to ensure (\ref{Equ: desired dual variables}).

\subsection{Auction Iteration}
\label{Subsection: Iterative Updating}

\begin{algorithm}[t]
	\caption{\textit{LtD Mechanism}}\label{Algorithm: A} 
	\SetKwInOut{Input}{Input}
	\SetKwInOut{Output}{Output}  
	\textbf{Initial} $\gamma>0$, $\epsilon>0$.	\\
	\For {$t=1$ \KwTo $T$ } 
	{
		\textbf{Initial} $k=0$ and $\bm{\lambda}^{t[1]}=(\lambda_{i,n}^{t[1]}:\forall n\in\mathcal{N},i\in\mathcal{I})$.\label{Line: Initialize}
		
		
		\Repeat{$\big|\bm{x}^{t[k]}-\bm{y}^{t[k]}\big| \le\gamma$ \label{Line: error}
		}
		{			
			\textbf{Set} $k=k+1$
			
			Broker announces allocation rule based on $\bm{\lambda}^{t[k]}$: 
			$X^{t}_{i,n}(\bidSP)
			={\bidSP}/{\lambda^{t[k]}_{i,n}}$ and  $Y^{t}_{i, n}(\bidPoI)
			={\lambda^{t[k]}_{i,n}}/{\bidPoI}$ \label{Line: allocation}
				
			Broker announces payment rule based on $\bm{\lambda}^{t[k]}$: 
			$\Pi^{t}_{n}(\bm{\bidSP}_{n})=\sum\limits_{i\in\mathcal{I}}s_{i,n}$ and $\Phi^{t}_{i}(\bm{\bidPoI}_{i})=\sum\limits_{n\in\mathcal{N}}\frac{(\lambda^{t[k]}_{i,n})^2}{\bidPoI_{i, n}}$
			\label{Line: pricing}
			
			Platform $n\in\mathcal{N}$ bids $\bm{\bidSP}^{t[k]}_{n}=(\bidSP_{i,n}^{t[k]}:\forall i\in\mathcal{I})$. \label{Line: bid}
			
			PoI $i\in\mathcal{I}$ bids $\bm{\bidPoI}^{t[k]}_{i}=(\bidPoI_{i,n}^{t[k]}:\forall n\in\mathcal{N})$.
			
			Broker determines uploading rates according to $x^{t[k]}_{i,n}
			=X^{t}_{i,n}(\bidSP^{t[k]}_{i,n})$ and $y^{t[k]}_{i, n}=Y^{t}_{i, n}(\bidPoI^{t[k]}_{i, n})$ \label{Line: determine_rate}
			
			Broker updates $\bm{\lambda}^{t[k]}$ according to \label{Line: update}
			\begin{equation}\label{Equ: adjust lambda}
				\textstyle
				\lambda^{t[k+1]}_{i,n} 
				= \Big[ \lambda^{t[k]}_{i,n} 
				+ \epsilon\Big( x^{t[k]}_{i,n}-y^{t[k]}_{i,n} \Big) \Big]^+,\ \forall n,i.
			\end{equation}
		}
		
		System runs with $(\bm{x}^{t[k]},\bm{y}^{t[k]})$. 
		Broker pays $\Phi_{i}^{t}(\bm{\bidPoI}_{i}^{t[k]})$ to PoI $i$ and charges $\Pi^{t}_{n}(\bm{\bidSP}_{n}^{t[k]})$ from platform $n$. \label{Line: final}
	}
\end{algorithm}

Recall that the target consistency price $\bm{\tilde{\lambda}}^{t}$ is characterized in (\ref{Equ: KKT original}) based on the private gradient information.
The broker cannot obtain $\bm{\tilde{\lambda}}^{t}$ directly, but can iteratively run Auction \ref{Auction: } and adjust the consistency price based on the intermediate outcomes.
Algorithm \ref{Algorithm: A} presents the LtD mechanism.
In each period $t$, the broker initializes the consistency price $\bm{\lambda}^{t[1]}$ (i.e., Line \ref{Line: Initialize}), and then repeats the following procedure until the 
termination criterion (i.e., Line \ref{Line: error}) holds.

\begin{itemize}		
	\item \textit{Lines \ref{Line: allocation} \& \ref{Line: pricing}:}
	The broker announces the allocation rule and the payment rule based on the consistency price  $\bm{\lambda}^{t[k]}$.
	
	\item \textit{Lines \ref{Line: bid}$\sim$\ref{Line: determine_rate}:}
	The platforms and PoIs submit the bids, and the broker calculates the uploading rate.
	
	\item \textit{Lines \ref{Line: update}:}
	The broker adjusts the consistency price according to (\ref{Equ: adjust lambda}) based on the parameter $\epsilon>0$.
\end{itemize}

If the consistency constraints hold within the error bound $\gamma$ (i.e., Line \ref{Line: error}), then the iteration in this period stops.
The system runs according to the final uploading rate, i.e., Line \ref{Line: final}.
The broker will charge the platforms and reimburse the PoIs accordingly.
Furthermore, Lemma \ref{Lemma: Convergence} presents the convergence of the auction iteration in Algorithm \ref{Algorithm: A}.
The proof relies on the concavity of social welfare and follows the rationale of the proof in Chapter 22 \cite{nisan2007algorithmic},

\begin{lemma}\label{Lemma: Convergence}
	In each period $t$, we have $\lim_{k\rightarrow\infty}
	\bm{\lambda}^{t[k]}
	=
	\bm{\tilde{\lambda}}^{t}$.
\end{lemma}

Lemma \ref{Lemma: Convergence} indicates that the consistency price sequence $\{\bm{\lambda}^{t[k]}:\forall k\ge1\}$ generated in Algorithm \ref{Algorithm: A} converges to  $\bm{\tilde{\lambda}}^{t}$ in each period $t$, which ensures (\ref{Equ: desired dual variables}) in Proposition \ref{Proposition: KKT equivalent}.
Section \ref{Subsection: One Period} will show that the auction iteration can quickly converge.

So far, we have completed the mechanism design guided by Proposition \ref{Proposition: KKT equivalent}.
We are ready to formally present the theoretical performance of the LtD mechanism.

\section{Performance Analysis}
\label{Section: Performance}
This section presents the theoretical performance of the LtD mechanism.
To start with, combining Proposition \ref{Proposition: KKT equivalent} and Lemmas \ref{Lemma: Reward PoI}$\sim$\ref{Lemma: Convergence}, we obtain the truthfulness and optimality results in Theorem \ref{Theorem: truthfulness} and Theorem \ref{Theorem: performance}, respectively.
\begin{theorem}[Truthfulness]\label{Theorem: truthfulness}
	The LtD mechanism can ensure the truthfully bidding of the platforms and PoIs.
\end{theorem}

\begin{theorem}[Optimality]\label{Theorem: performance}
	The LtD mechanism achieves the same performance as $\{\bm{\tilde{x}}^{t}:\forall t\in\mathcal{T}\}$ defined in (\ref{Equ: maximize virtual social welfare}).
\end{theorem}

Next we introduce the benefits of the broker, PoIs, and platforms under LtD mechanism.

\begin{theorem}[Budget Balance]\label{Theorem: budget balance}
	Under the LtD mechanism, the total payment of the platforms equals to the total reimbursement to the PoIs in each period.
\end{theorem}

Theorem \ref{Theorem: budget balance} indicates that there is no need for the broker to inject or take money when running the LtD mechanism, thus the broker always maintains a strictly balanced budget in each period.
This is what a non-profit broker (e.g., the government) desires in practice.

\begin{theorem}[Voluntary Participation]
	\label{Theorem: IR for PoI}
	Under the LtD mechanism, each PoI $i$ achieves a non-negative payoff in each period.
\end{theorem}

Theorem \ref{Theorem: IR for PoI} shows that the LtD mechanism ensures the \textit{voluntary participation} for each PoI.
That is, the PoI will voluntarily upload its real-time status, which helps attract more PoIs to join the status acquisition system.

\section{Numerical Results}
\label{Section: Numerical results}
This section provides the numerical results and evaluates the LtD mechanism.
To proceed the evaluation, we consider that each platform $n$ has the following $\alpha$-fair utility \cite{mo2000fair}
\begin{equation}\label{Equ: numerical_utility}
	U^{t}_{n}(\bm{x}_{n})
	=
	\sum\limits_{i\in\mathcal{I}} 
	\rdmSP^{t}_{i,n}\cdot\frac{x_{i,n}^{1-\alpha}}{1-\alpha},
	\quad\forall n\in\mathcal{N},
\end{equation}
where $\alpha\in(0,1)$ is the coefficient of relative risk aversion and $\rdmSP^{t}_{i,n}>0$ captures how platform $n$ values the status acquisition from PoI $i$.
Moreover, we consider the following PoI cost
\begin{equation}\label{Equ: numerical_cost}
	C^{t}_{i}(\bm{x}_{i})
	=
	\sum\limits_{n\in\mathcal{N}}l_{i, n}^{t}x_{i,n}+\pi_{i}^{t}e_{i}^{t}x_{i,n}^{2} ,
	\quad\forall i\in\mathcal{I},
\end{equation}
where $l_{i, n}^{t}$ is the sensitivity of PoI $i$'s privacy loss with respect to platform $n$.
Moreover, $\pi^{t}_{i}$ and $e^{t}_{i}$ correspond to the electricity price and the energy consumption level, respectively.

We first demonstrate the per-period convergence result in Section \ref{Subsection: One Period}.
We then evaluate the long-term performance of the LtD mechanism in Section \ref{Subsection: Multi-Period}.

\subsection{One-Period Convergence}
\label{Subsection: One Period}
We consider a small scenario with a platform and two PoIs to shed light on the auction iteration within a period.
According to (\ref{Equ: virtual SW}), the value $Q^{t}_{n}/\V$ affects the uploading rate in period $t$.
Specifically, $\V$ also affects the virtual queue $\{Q^{t}_{n}:\forall t\in\mathcal{T}\}$ over multiple periods.
To illustrate the convergence results, we will view $Q^{t}_{n}/\V$ as a parameter and investigate its impact within a specific period.
We will work on the multi-period evaluation in Section \ref{Subsection: Multi-Period}.

\begin{figure}
	\setlength{\abovecaptionskip}{0pt}
	\setlength{\belowcaptionskip}{0pt}
	\centering
	\subfigure[${Q^{t}_{n}}/{\V}=0.1$]
	{\label{fig: Convergence_1}\includegraphics[height=0.375\linewidth]{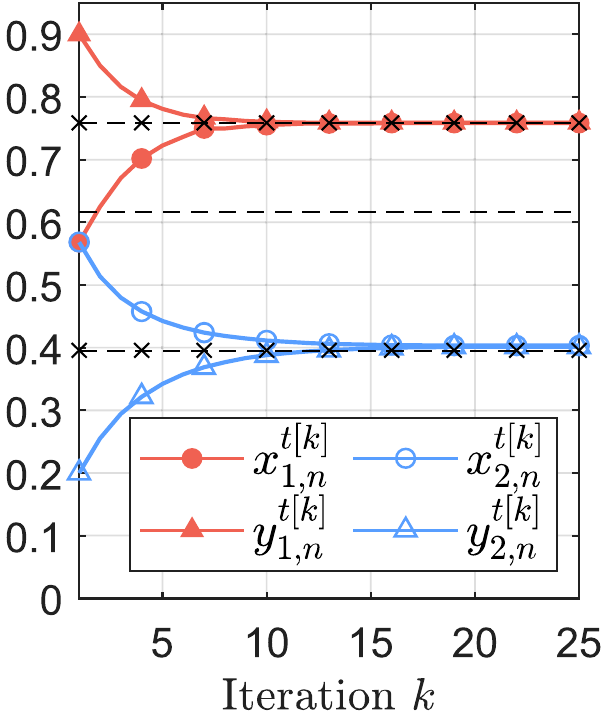}}\ 
	\subfigure[${Q^{t}_{n}}/{\V}=1$]
	{\label{fig: Convergence_2}\includegraphics[height=0.375\linewidth]{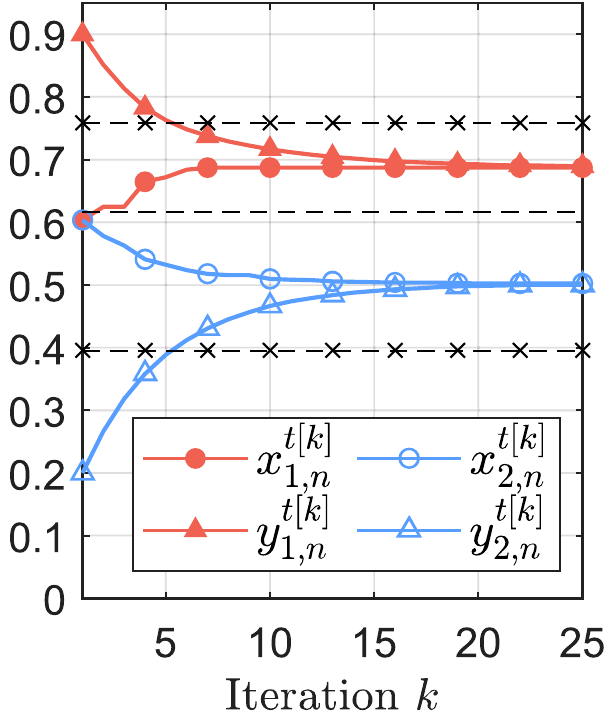}}\ 
	\subfigure[${Q^{t}_{n}}/{\V}=100$]
	{\label{fig: Convergence_3}\includegraphics[height=0.375\linewidth]{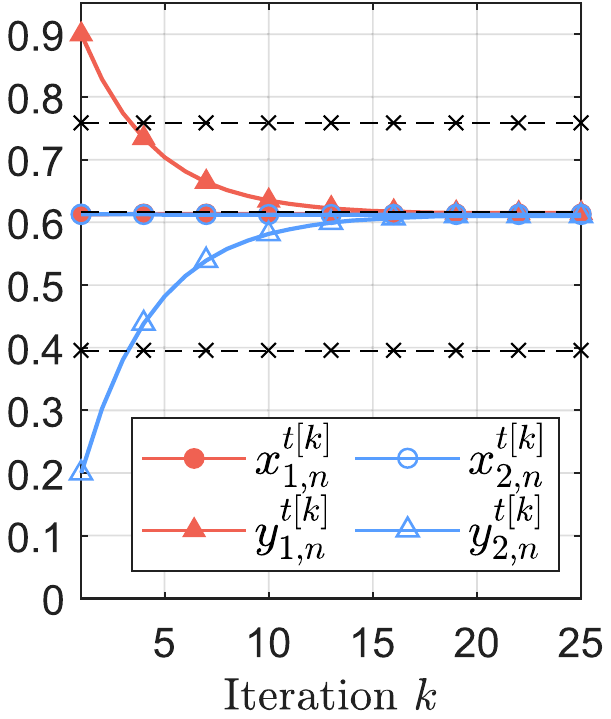}}	
	\caption{Illustration of convergence.\vspace{-5pt}}
	\label{fig: Convergence}
\end{figure}

Fig. \ref{fig: Convergence} plots the convergence results under different values of $Q^{t}_{n}/\V$.
In each sub-figure, the horizontal axis represents the iteration index $k$.
The two red curves plot the uploading rates related to PoI 1.
The two blue curves represent the uploading rates related to PoI 2.
Moreover, the two cross lines denote the social optimal uploading rates given the realized random factors.
The dash line (without marker) represents the age-minimizing uploading rate.
The step-size parameter is $\epsilon=0.1$.
We have two-fold observations based on Fig. \ref{fig: Convergence}.
\begin{itemize}
	\item In each sub-figure, the uploading rates will converge within twenty iterations, which implies that LtD mechanism is efficient to implement in practice.
	
	\item When $Q^{t}_{n}/\V$ is small, e.g., Fig. \ref{fig: Convergence_1}, the uploading rates almost converge to the social optimal results.
	This is because that the real social welfare in (\ref{Equ: virtual SW}) dominates the virtual social welfare in this case.
	In contrast, when $Q^{t}_{n}/\V$ is large, e.g., Fig. \ref{fig: Convergence_3}, the uploading rates converge to the age-minimizing results.
\end{itemize}

Next we move on to the multi-period evaluation and investigate the long-term impact of $\V$ under the LtD mechanism.

\begin{figure}
	\centering
	\begin{minipage}{0.49\linewidth}
		\setlength{\abovecaptionskip}{5pt}
		\setlength{\belowcaptionskip}{0pt}
		\centering
		\includegraphics[height=0.64\linewidth]{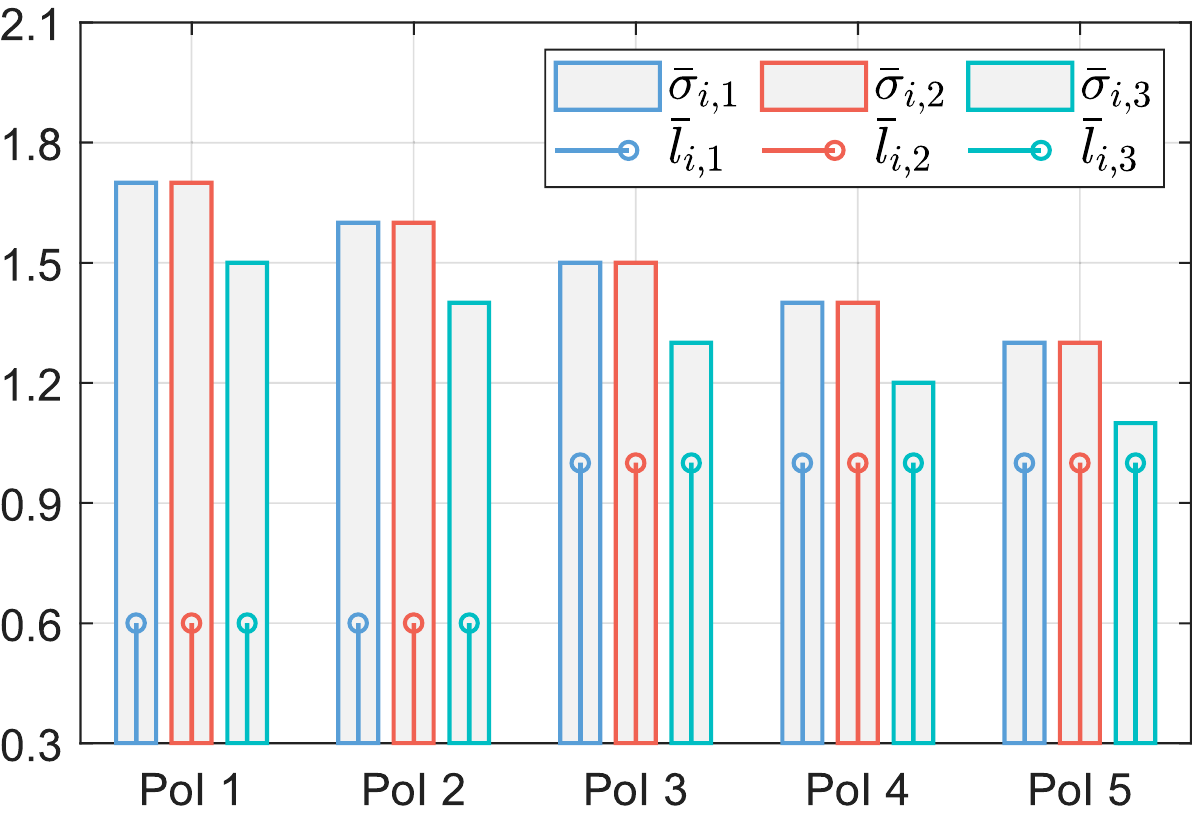}
		\caption{Scenario}
		\label{fig: Scenario}
	\end{minipage}
	\begin{minipage}{0.49\linewidth}
		\setlength{\abovecaptionskip}{3pt}
		\setlength{\belowcaptionskip}{0pt}
		\centering
		\includegraphics[height=0.66\linewidth]{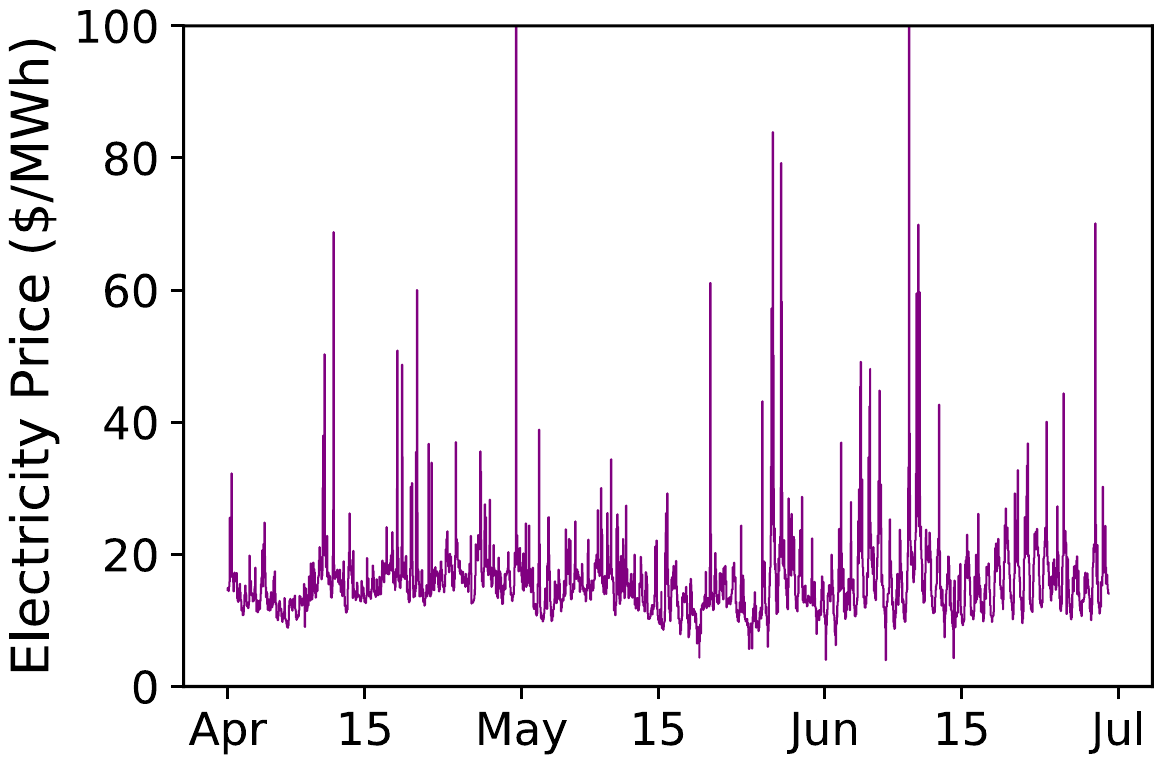}
		\caption{Electricity price}
		\label{fig: price}
	\end{minipage}
\end{figure}

\subsection{Multi-Period Evaluation}
\label{Subsection: Multi-Period}
\begin{figure}
	\setlength{\abovecaptionskip}{0pt}
	\setlength{\belowcaptionskip}{0pt}
	\centering
	\subfigure[Social welfare]
	{\label{fig: SW}
		\includegraphics[height=0.33\linewidth]{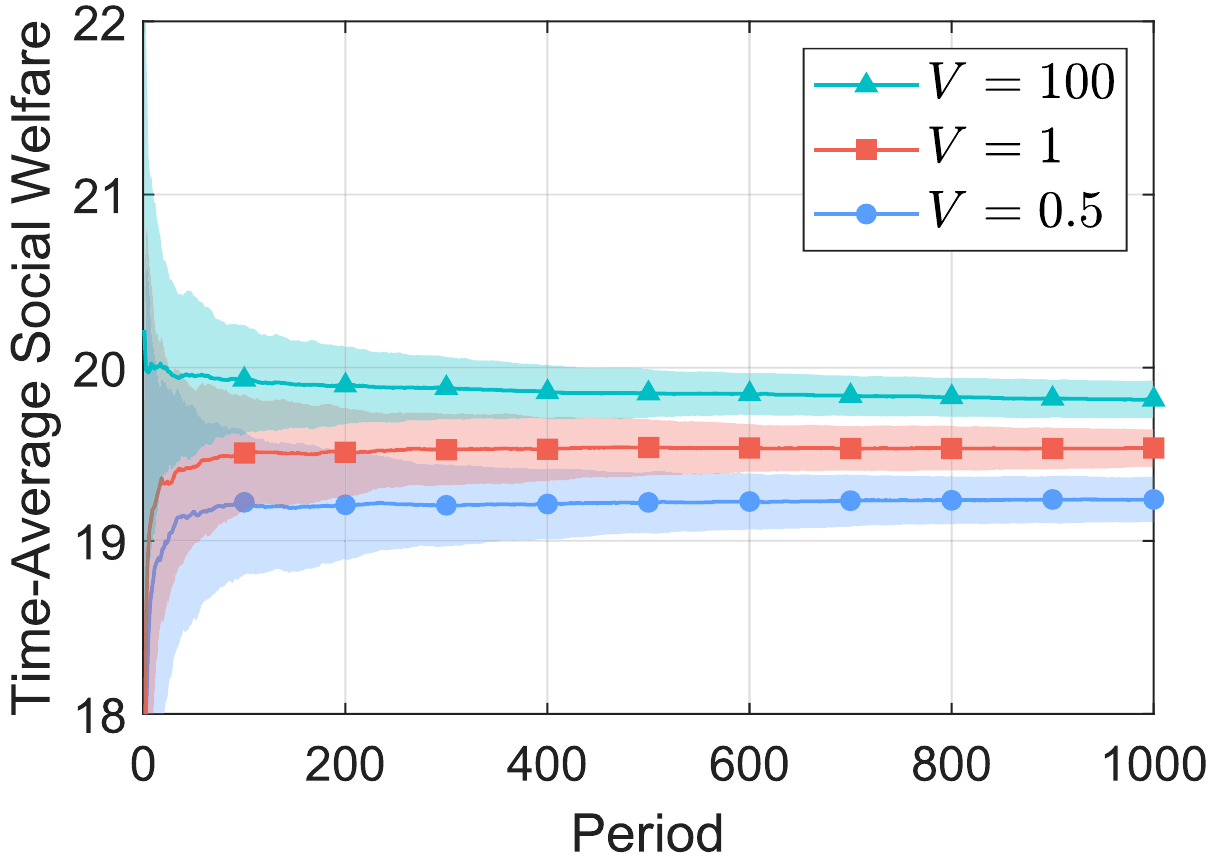}}\ 
	\subfigure[Platform age]
	{\label{fig: SP_AoI}
		\includegraphics[height=0.33\linewidth]{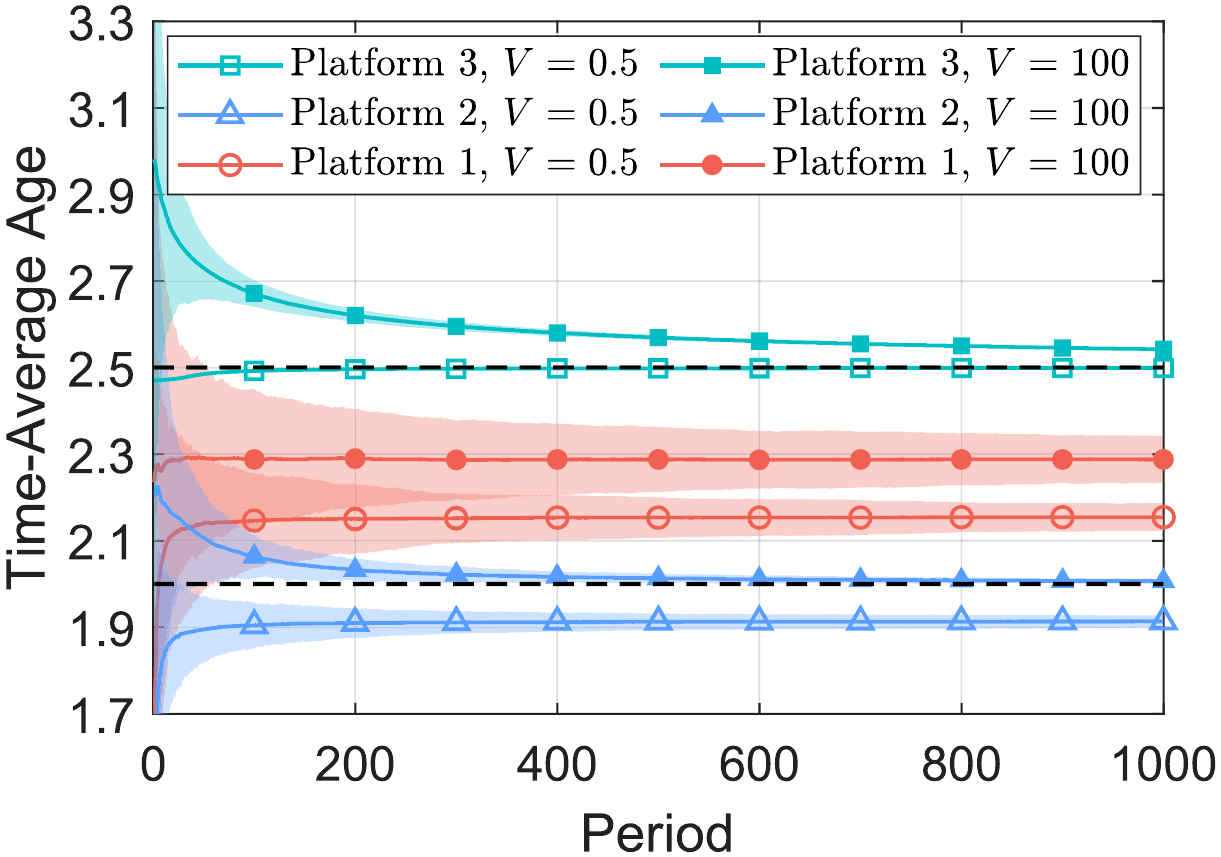}}	
	\caption{Performance of LtD mechanism}
	\label{fig: performance}
\end{figure}

We evaluate the long-term performance of LtD mechanism and consider three platforms and five PoIs as shown in Fig. \ref{fig: example}.
Specifically, we randomly generate the valuation $\rdmSP_{i,n}^{t}$ and privacy loss $l_{i,n}^{t}$ according to the truncated normal distribution with means $\bar{\rdmSP}_{i,n}$ and $\bar{l}_{i,n}$ shown in Fig. \ref{fig: Scenario}, respectively.
Note that platform 1 and platform 2 have the same average valuation, which is higher than platform 3.
We specify the computing capability and freshness threshold according to $\{\proc_{1},\proc_{2},\proc_{3}\}=\{10,10,5\}$ and $\{\AoIthr_{1},\AoIthr_{2},\AoIthr_{3}\}=\{2.5,2,2.5\}$, respectively.
That is, platform 1 and platform 2 have the advantage in computing capability, while platform 2 also has a more strict freshness requirement.
To quantify the energy expenditure, we use the real-world electricity market price in US \cite{PJM}.
Fig. \ref{fig: price} shows the hourly price from April to June in 2020.
We evaluate the LtD mechanism for one hundred times and plot the results in Fig. \ref{fig: performance}.
\begin{itemize}
	\item Fig. \ref{fig: SW} shows the time-average social welfare, where the three curves correspond to $\V\in\{0.5,1,100\}$.
	Specifically, a large $\V$ increases the social performance, but also affects the platform freshness as shown in Fig. \ref{fig: SP_AoI}.
	
	\item Fig. \ref{fig: SP_AoI} plots the time-average platform ages under $\V\in\{0.5,100\}$.
	The two dash lines represent the freshness thresholds.
	Comparing the two square curves (or triangle curves) shows that a large $\V$ increases the number of periods to satisfy the freshness conditions.
\end{itemize}

\begin{figure}
	\setlength{\abovecaptionskip}{0pt}
	\setlength{\belowcaptionskip}{0pt}
	\centering
	\subfigure[Platform payoff]
	{\label{fig: Payoff_SP}
		\includegraphics[height=0.345\linewidth]{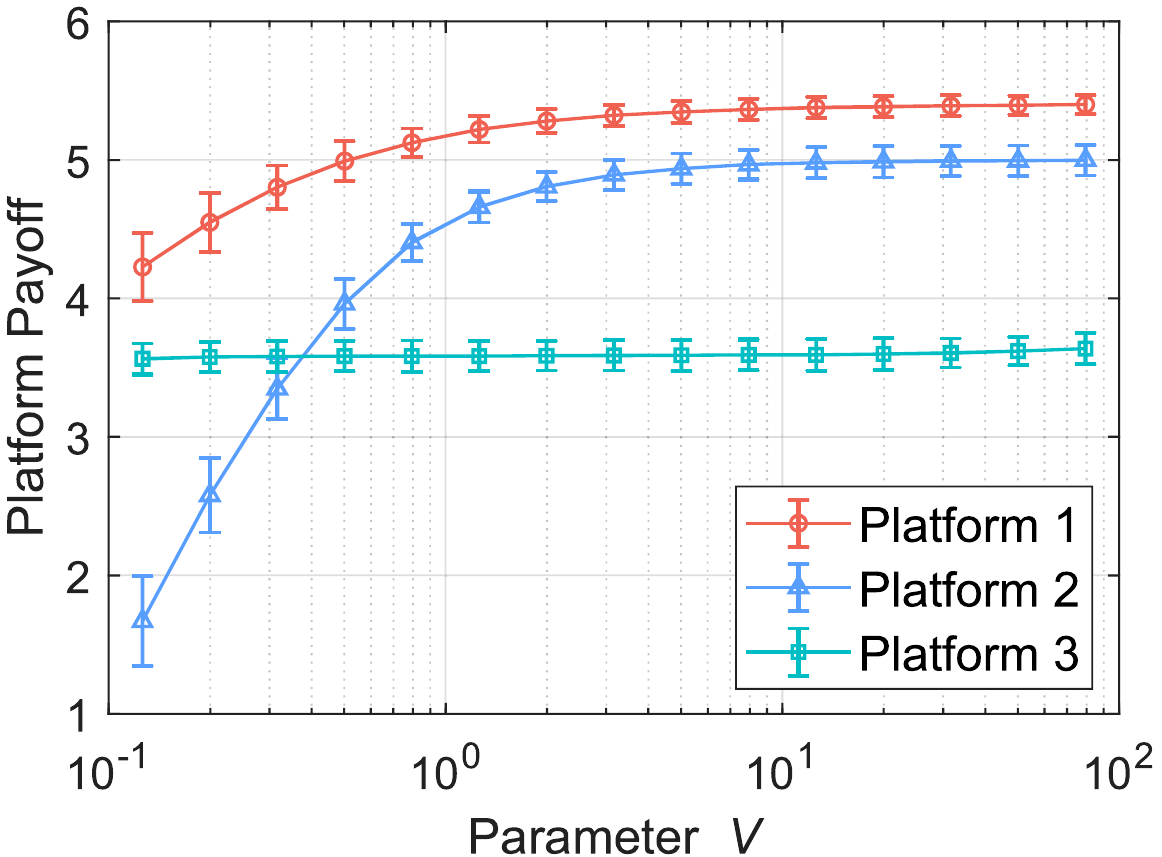}}
	\subfigure[PoI payoff]
	{\label{fig: Payoff_PoI}
		\includegraphics[height=0.345\linewidth]{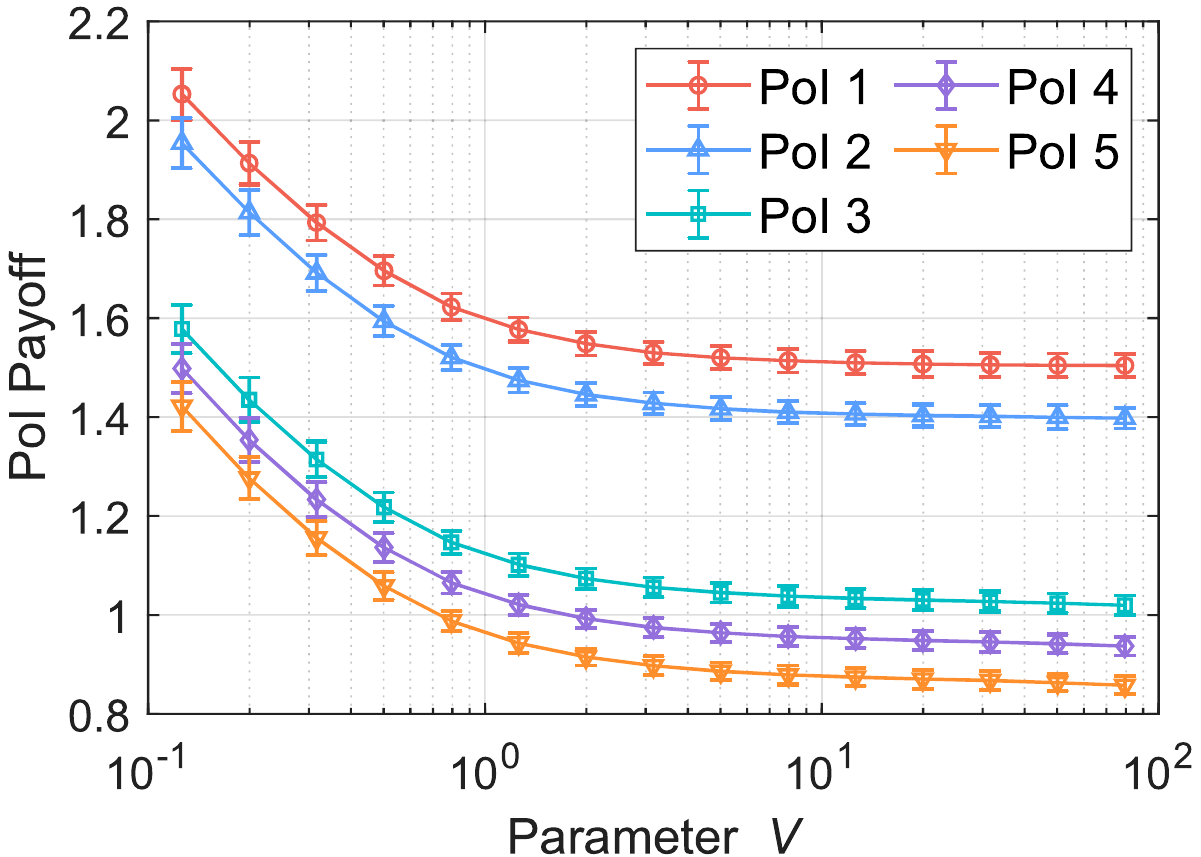}}	
	\caption{Impact of $V$ on the payoffs of platforms and PoIs}
	\label{fig: V}
\end{figure}

Fig. \ref{fig: V} investigates the impact of the parameter $\V$ on the long-term payoffs of the platforms and PoIs under LtD mechanism.
Overall, the payoff of each platform increases in $\V$, since a larger $\V$ means more attention on payoff maximization in the platform's long-term bidding problem.
Fig. \ref{fig: Payoff_PoI} shows that the payoff of each PoI decreases in $\V$ and converges to non-negative values, which verifies the voluntary participation property for each PoI in Theorem \ref{Theorem: IR for PoI}.

\section{Conclusion and Future Work}
\label{Section: Conclusion}
In this paper, we investigate the real-time operation of a status acquisition system with multiple self-interested platforms and PoIs.
The platforms and PoIs have different objectives, which are private and time-varying.
To resolve the time-varying information asymmetry, we devise a long-term decomposition (LtD) mechanism, which helps the market broker manipulate the interactions between the PoIs and platforms.
We show that the LtD mechanism retains the same performance compared to the symmetric information scenario, and asymptotically ensures the platform freshness conditions. 

In the future, we would like to extend the results in this paper from the following aspects.
First, it is interesting to compare different platform updating disciplines (e.g., \textit{first-come-first-update} and \textit{last-come-first-update}).
Second, it is also interesting to consider the status correlation among PoIs.
%

\bibliographystyle{IEEEtran}
\bibliography{ref}

\end{document}